\documentclass[%
 aip,
 rsi,
 amsmath,amssymb,
 reprint,%
]{revtex4-1}

\usepackage{graphicx}
\usepackage{dcolumn}
\usepackage{bm}

\usepackage[utf8]{inputenc}
\usepackage[T1]{fontenc}
\usepackage{mathptmx}
\usepackage{etoolbox}
\usepackage[separate-uncertainty=true]{siunitx}
\usepackage{comment}
\usepackage[dvipsnames]{xcolor}
\usepackage{upgreek} 
\usepackage[caption=false]{subfig}
\usepackage{paralist}
\usepackage{hyperref}
\usepackage{ulem}
\newcommand{\di}{\operatorname{d}\!}
\newcommand\Tstrut{\rule{0pt}{2.6ex}}         
\newcommand\Bstrut{\rule[-0.9ex]{0pt}{0pt}}   

\makeatother
\begin{document}

\preprint{AIP/123-QED}

\title{Enhancing Neutron Measurement Accuracy with Bubble Detectors at Laser-Driven Neutron Sources}
\author{Stefan Scheuren}
  \email{sscheuren@ikp.tu-darmstadt.de}
\affiliation{ 
Technische Universität Darmstadt,
Institut für Kernphysik (IKP), Schlossgartenstr. 9, Darmstadt, 64289, Germany
}%

\author{M. \'{A}ngeles Mill\'{a}n-Callado}
\affiliation{Universidad de Sevilla, Dept. de Física Atómica, Molecular y Nuclear, Avda. Reina Mercedes s/n, Seville, 41012, Spain}
\affiliation{Centro Nacional de Aceleradores (Universidad de Sevilla-Junta de Andalucía-CSIC),  C/ Thomas Alva Edison 7, Seville, 41092, Spain}

\author{Jonas Kohl}%

\author{Tim T. Jäger}%

\author{Markus Roth}
\affiliation{ 
Technische Universität Darmstadt,
Institut für Kernphysik (IKP), Schlossgartenstr. 9, Darmstadt, 64289, Germany
}%

\author{Christian Rödel}
\affiliation{University of Applied Science Schmalkalden, Faculty of Mechanical Engineering, Blechhammer 6-9, Schmalkalden, 98574,Germany}

\author{Aaron Alejo}
\affiliation{Instituto Galego de Física de Altas Enerxías, Universidade de Santiago de Compostela, Santiago de Compostela, 15705, Spain}

\author{Florian Kroll}

\author{Karl Zeil}

\author{Arnd Junghans}

\affiliation{
Helmholtz-Zentrum Dresden-Rossendorf
Dresden, 01328, Germany
}%

\author{Carlos Guerrero}
\affiliation{Universidad de Sevilla, Dept. de Física Atómica, Molecular y Nuclear, Avda. Reina Mercedes s/n, Seville, 41012, Spain}
\affiliation{Centro Nacional de Aceleradores (Universidad de Sevilla-Junta de Andalucía-CSIC),  C/ Thomas Alva Edison 7, Seville, 41092, Spain}

\author{Benedikt Schmitz}%
  \email{bschmitz@ikp.tu-darmstadt.de}
\affiliation{ 
Technische Universität Darmstadt,
Institut für Kernphysik (IKP), Schlossgartenstr. 9, Darmstadt, 64289, Germany
}%
\affiliation{Hochschule Darmstadt, Elektrotechnik und Informationstechnik (EIT), Birkenweg 8, 64295 Darmstadt}
\affiliation{European University of Technology, European Union}

\date{\today}

\begin{abstract}
Bubble detectors are widely used to measure neutron flux from laser-driven sources employing a pitcher-catcher setup, due to their insensitivity to intense $\gamma$-ray backgrounds and strong electromagnetic pulses (EMP).\\
This paper presents a method to account for the neutron energy-dependent response of bubble detectors, enabling accurate conversion of bubble counts into neutron flux at the detector location.
The proposed method is based on the accurate reconstruction of the response function using a surrogate model. The resulting model is convoluted with the (normalized) expected/measured neutron spectrum to obtain an effective measure of the bubble detector's response, herein referred to as effective $c$ or $c_\text{eff}$. This effective value for the response is energy-independent after the convolution. In this way, our approach includes the spectral distribution of neutrons arriving at the detector to determine the integral neutron flux. Analyzing our experimental results obtained at the DRACO PW laser and comparing the results to previously used methods to obtain neutron fluxes from bubble detectors returns a reduction in neutron flux of up to \SI{31}{\%}. Results from the method detailed in this paper agree with in-depth experimental setup Monte Carlo simulations, with deviations of less than \SI{10}{\%}. We furthermore discuss the inherent limitations of our method with regard to its uncertainty and highlight the influence of neutron scattering in bubble detector measurements. For our experimental setup at the DRACO laser, up to \SI{47}{\%} of the detected neutrons arrive at the detector after undergoing at least one scattering event.
\end{abstract}
\maketitle

\section{Introduction}\label{sec:intro}
    Laser-driven neutron sources (LDNS) have made great strides forward in recent years \cite{alvarez2014laser, alejo2016recent, yogo2023advances} with first demonstrations of non-destructive sample analysis, employing different neutron measurement techniques \cite{zimmer2022demonstration, yogo2023laser}. LDNS commonly use the pitcher-catcher scheme \cite{lancaster2004characterization}, in which protons are accelerated by directing an ultrahigh-intensity laser pulse onto a laser target, usually a thin (nm to $\upmu$m thickness) metal or plastic foil, the so-called pitcher. In this way, protons can be accelerated to several tens of \si{MeV} through the so-called target normal sheath acceleration \cite{snavely2000intense, wilks2001energetic} (TNSA) mechanism. It needs to be mentioned that TNSA does not generate a pure proton beam. Instead, due to the complex acceleration mechanics, other ion species can also be accelerated, and MeV electrons and photons accompany the ion beam \cite{passoni2010target, mckenna2013laser}.
    Subsequently, the ions are directed onto a suitable converter material \cite{kar2016beamed} (the catcher), often beryllium \cite{roth2013bright, Kleinschmidt2018, yogo2023laser}, copper \cite{roth2013bright, pomerantz2014ultrashort}, or lithium-fluoride \cite{lancaster2004characterization, zimmer2022demonstration}, where neutrons are generated via nuclear reactions.
    Due to their unique characteristics, such as small form factor, high peak flux, and short primary neutron pulse duration, these sources have been of immense interest as a promising alternative to high-intensity conventional accelerator-based neutron sources \cite{alvarez2014laser, alejo2016recent, yogo2023advances}.

    One drawback of LDNS is the high noise environment they generate, such as a prompt electromagnetic pulse (EMP) interfering with detection and data acquisition systems \cite{bradford2018emp, seimetz2020electromagnetic} or the high x-ray/$\gamma$-ray background, as many different interaction processes take place within the laser pulse duration (typically $\lesssim \SI{1}{ps}$). 
    Also important to note is that due to its short acceleration time all of the generated neutrons come at once generating a high instantaneous neutron flux and results in a pile up inside of electronic detectors.
    Since the current repetition rate of laser systems is relatively low, only single shot diagnostics are necessary to be used so far. 
    Due to this high noise environment interfering with the measurement, it remains a challenge to determine absolute neutron numbers accurately at these sources within reasonable time frames. 
    To circumvent these limitations, the community has adopted Bubble Detectors \cite{Ing1984,Ing1997} (BDs) manufactured by Bubble Technology Industries (BTI) to measure the integral fast neutron flux per solid angle. 
    The radiation field outside of the vacuum chamber only consists of neutron and gamma signals, while ions and charges particles are stopped in the chamber walls.
    The BDs' are insensitive to gamma radiation, because they rely on a kinematic detection approach. 
    In short, BDs are filled with a polymer gel which contains super-heated halogen and/or hydrocarbon droplets. 
    When a neutron interacts with a BD, energy is deposited inside the gel. 
    The droplet vaporizes if the deposited energy exceeds a threshold, leaving a visible bubble behind. 
    The energy threshold is dependent on the manufacturing process and can range from \SIrange{10}{10000}{\kilo\electronvolt}\cite{BTI}.
    Gammas do not deposit as much energy inside the gel, such that they can be assumed to be insensitive to gammas\cite{Smith2014,BTI}.
    The generated bubbles can therefore be counted and translated into a neutron dose. 
    Their original purpose was to be used as a personal neutron dosimeter. 
    However, the kinetic approach to detection makes the detectors highly suitable for the environment present at laser-driven neutron sources. 
    Therefore, bubble detectors have emerged as on of the most used diagnostic for measuring and comparing neutron yields at LDNS \cite{mirfayzi2015calibration, yogo2023advances}.
    They are insensitive to these sources' harsh $\gamma$ noise environment and their dominant EMP, as their deposited energy is too low to trigger the bubble formation inside the contained superheated liquid.

    To calculate the neutron number $\hat{N}$ from a count of bubbles $b$, \autoref{eq:bubbles} is used \cite{Kleinschmidt2018}
    \begin{align}
        \hat{N} = \frac{b\cdot d^2}{\hat{c}_\text{avg} \cdot s_0}, \text{ with } \left[\hat{N}\right] = \frac{1}{\text{sr}},
    \label{eq:bubbles}
    \end{align}
    where $d$ is the distance from the source to the detector in cm, $s_0$ is the sensitivity of the bubble detector in bubbles per mrem and $\hat{c}_\text{avg}$ is based on the neutron energy-dependent response function $\hat{c}(E_\text{n})$ of the bubble detector. However, since BDs measure the integral neutron flux, the energy distribution of the neutrons at the position of the detector is unknown. For this reason $\hat{c}(E_\text{n})$ is commonly approximated by averaging the response function over the energy region of interest, with often-used values ranging from $\hat{c}_\text{avg} = \SIrange[range-units=single]{3.0}{3.5}{bubbles \cdot cm^2 /neutron}$ \cite{pomerantz2014ultrashort, Kleinschmidt2018, trefferthigh} for \SI{1}{bubble} per mrem \cite{jung2013characterization}. This approach can result in the incorrect estimation of neutron numbers, as the calculation neglects the influence of the spectral distribution of the generated neutron beam. 
    As shown later, the amount of scattered neutrons is also to be considered since this leads to multiple counts of the same neutrons with different bubble counts. 
    

    In this work, we present a method to account for the energy dependence of the response functions, similar to the one used by Jung \textit{et al.} \cite{jung2013characterization}. The method introduces an experiment-specific, effective value for $\hat{c}(E_\text{n})$, which we will refer to as effective $c$ or $c_\text{eff}$. This $c_\text{eff}$ is given for \SI{1}{bubble} per mrem as is in \cite{Ing1997, jung2013characterization, Kleinschmidt2018}. We use the calculated value for $c_\text{eff}$ to determine the neutron flux and its uncertainty for experimental data recorded during an experimental campaign at the DRACO laser system at the HZDR in Dresden, Germany\cite{Ziegler_2021}.
    Furthermore, we will compare the neutron fluxes calculated by our proposed method to values calculated using previously published approaches \cite{jung2013characterization, Kleinschmidt2018}. Additionally, we conducted in-depth Monte Carlo simulations of the experimental setup, which we will use as a point of comparison to the fluxes obtained from the bubble detector measurements, utilizing the different methods.
    
    Finally, during the discussion, we will highlight general problems arising from using BDs to compare different laser facilities by highlighting the contribution of neutron scattering. In conclusion, we provide recommendations for using BDs at LDNS to ensure better comparability between different setups. 
    
    Our investigation uses Monte Carlo simulations conducted with PHITS \cite{sato2024recent}, a nuclear Monte Carlo code developed and maintained by the Japan Atomic Energy Agency (JAEA). PHITS has been thoroughly benchmarked \cite{iwamoto2022benchmark} for different applications. A comparison between PHITS and experimental measurements about neutron generation can be found in \cite{iwamoto2017benchmark, iwamoto2023benchmark}, where good agreement was observed. It should be noted up front that the approach presented in this paper depends strongly on the accuracy of the PHITS simulations.
    
    Our developed code and surrogate model for analysing the bubble detector data is openly available on Gitlab under the Creative Commons Attribution 4.0 license.

\section{Proposed methodology\label{sec:Methods}}
    This work's approach to obtaining the neutron flux from bubble detector data is based on the accurate reconstruction of the bubble detector neutron energy-dependent response function, published by the manufacturer (BTI), using a surrogate model. The model's output is then convoluted with the neutron energy-dependent $E_\text{n}$, differential neutron spectrum to calculate a setup-specific effective value of the response $c_\text{eff}$. This calculation will be discussed in the manuscript in \autoref{sec:cdetermination}, as it has the most influential effect on calculating neutron numbers. 
    
    Additionally, we introduce corrections to the sensitivity to obtain an effective sensitivity $s_\text{eff}$, which takes the temperature $T$ of the measurement environment, as well as bubble detector production variances, into account. Determination of $s_\text{eff}$ is discussed in the appendix \ref{app:s_eff}, as it has a reduced influence on the overall results compared to $c_\text{eff}$.
    From these effective values, the neutron number $N$ can be calculated as in \autoref{eq:bubbles}
    \begin{align}
        N = \frac{b\cdot d^2}{c_\text{eff} \cdot s_\text{eff}}.
    \label{eq:adj-bubbles}
    \end{align}
    Please note that $N$ and $\hat{N}$ are different; $ N$ is the number of neutrons this method yields, while $\hat{N}$ is the number of neutrons yielded if the response function is approximated as a constant.
    The uncertainty corresponding to \autoref{eq:adj-bubbles} is calculated via Gaussian error propagation and is given by 
    \begin{align}
        \Delta N = \sqrt{\sum_i \left(\frac{\partial N}{\partial x_i}\cdot \Delta x_i\right)^2},
        \label{eq:uncert-partial}
    \end{align}
    
    where $x_i \in \left[ b, d, c_\text{eff}, s_\text{eff}, E_\text{n}, T \right]$ are the variables of \autoref{eq:adj-bubbles} and $\Delta x_i \in \left[ \Delta b, \Delta d, \Delta c_\text{eff}, \Delta s_\text{eff}, \Delta T, \Delta E_n \right]$ the uncertainty of the associated variables. A full expression of \autoref{eq:uncert-partial} can be found in the appendix \ref{app:uncert}. 

    In the first step, we need to determine  $c_\text{eff}$. For this, we reconstruct and fit the response function based on measured and simulated data published by BTI \cite{Ing1997, Smith2014} and apply Bootstrapping and Gaussian Process Regression (\autoref{sec:cdetermination}). Afterwards, the reconstructed response function is convoluted with the expected (normalized) neutron spectrum. 
    The shape can be obtained from experimental data or, as in this study, from Monte Carlo simulations, see \autoref{sec:c_eff}. 
    We calculate $c_\text{eff}$ for different converter materials, based on otherwise identical inputs in \autoref{sec:c_eff}. 

    \subsection{Determination of an energy-dependent model for \textit{c} and its derivative \label{sec:cdetermination}}
\begin{figure*}
	   	\centering
		\subfloat[Model and plot with data and $2\sigma$ uncertainty interval.\label{fig:ModelWithBootstrap}]{\includegraphics[width=0.49\textwidth]{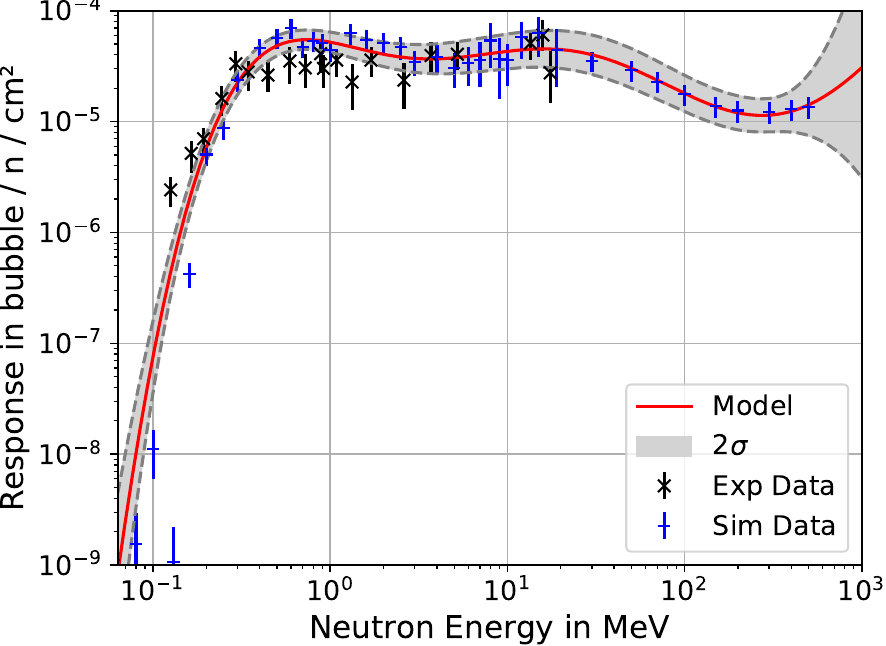}}\hfill
		\subfloat[Model and plot of the bootstrapped data. \label{fig:bootstrap}]{\includegraphics[width=0.49\textwidth]{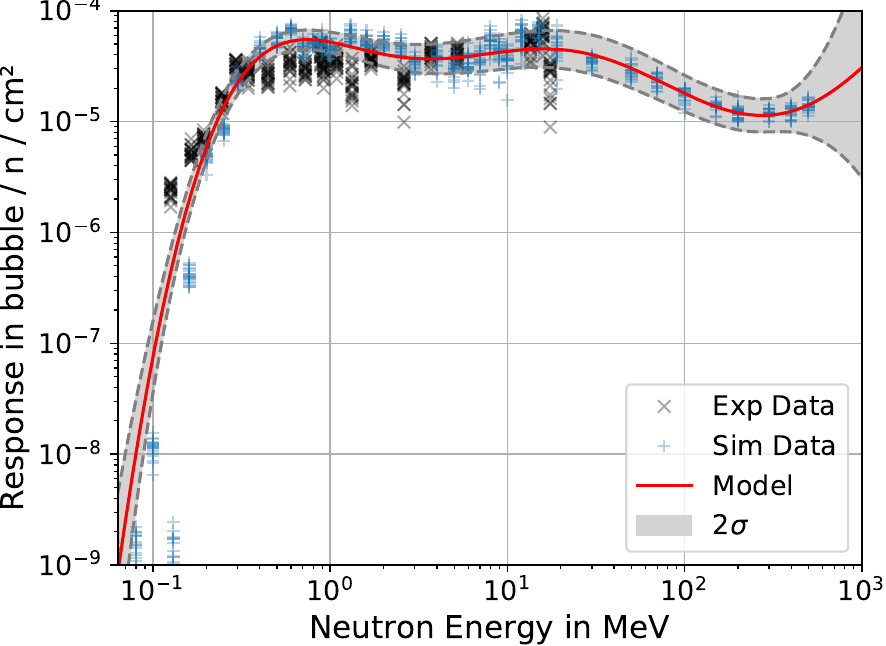}}
	   	\caption{Plot of the data and the model. (a) the given data, (b) the resampled data. The markers represent the measured and simulated bubble detector response to neutrons at a given energy, supplied by BTI \cite{Ing1997, Smith2014}. The red solid line is the reconstructed surrogate model, while the gray area with its dashed boundary indicates the model's $2\sigma$ uncertainty. The evaluation was done with $n=20$, where $n$ is the number of resamplings done. }
   \end{figure*}

    Looking at the response function in \autoref{fig:bootstrap} it can be seen that its overall form is complex, covering five orders of magnitude.
    For neutrons with energies around \SI{100}{keV} the response is very low, it then rises rapidly by up to four orders of magnitude and peaks around \SIrange[range-units=single]{500}{700}{keV}, before decreasing again by about a factor of two. It then rises again for neutrons with an energy of \SI{3}{MeV} or more. Thus, due to the shape of the response function, bubble detectors are sensitive to variations in the neutron spectrum present at the detector position. For example, a setup emitting most neutrons between \SIrange[range-units=single]{0.4}{1}{MeV} would generate more bubbles than a setup mainly emitting \SIrange[range-units=single]{2}{3}{MeV} neutrons.

    As a result of the complex nature of the detector response and its extensive dynamic range, a polynomial fit cannot be used to obtain a sufficient representation of the response function. Instead, we use a surrogate model to reproduce its features more accurately, the results of which are displayed by the red line in \autoref{fig:bootstrap}. To obtain and train the surrogate model, we use the SMT Toolbox\cite{Saves2024}.
    Reconstruction of the response function is based on measured and (Monte Carlo) simulated response data which BTI provided. The provided data on which the reconstruction is based were already published\cite{Ing1997, Smith2014}. The two data sets are joined together to form a more extensive set for the detector response, ranging from around \SI{80}{keV} to \SI{600}{MeV} and is indicated by the markers in \autoref{fig:bootstrap}.

    We fit a Gaussian process regression to the response data to account for the limited data and its distribution, see \autoref{fig:bootstrap}. 
    Gaussian process regression is based on model estimation using Bayesian optimization to a prior given by the Gaussian process. 
    Due to this, the model does not behave linearly, resulting in different curve behavior when the prior is updated\cite{RasmussenW06}.
    
    Both data sets have a finite accuracy; during model fitting, a bootstrapping method is applied in which we resample the value of the response function at a given energy within its associated uncertainty. The resampled values are randomly taken from the associated uncertainty interval and sampled according to the underlying Gaussian distribution. This is again done using the functions provided by the SMT Toolbox. 

    As is visible from the data points in \autoref{fig:bootstrap}, the simulated response data (blue + markers) deviated significantly from the measured response (black crosses) by up to a factor of 3 between \SIrange[range-units=single]{0.4}{1}{MeV}. During the model fitting, the simulation data in this energy range has an increased weight on the resulting model due to its smaller uncertainty (relative to the experimental data). This results in a biasing towards the simulated response function. Furthermore, since the simulation uncertainty is only based on the statistical uncertainty, it can become arbitrarily small if the number of simulated particles is sufficiently large.
    If simulation data besides the one taken from the work of Smith \textit{et al.}\cite{Smith2014} is taken into account, then this variance can be tailored to improve the model.
    
    Therefore, to counteract the biasing towards simulation data and to account for systematic uncertainties in the simulation, we studied the model output when applying a relative (meaning percentage-based) increase to the uncertainty of all simulated data points. This relative increase, given in percent of the data point's value, is referred to as $\sigma_\text{add}$ by us and allows us to adjust the weight that the simulated data carries (compared to the experimental data) when the model is trained. 
    More details and a comparison of the influence of $\sigma_\text{add}$ on the results can be found in \autoref{app:systematicuncertainty}.
    The final surrogate model is then trained on this modified data set, with $\sigma_\text{add}=\SI{10}{\%}$ and is displayed in \autoref{fig:bootstrap}.
    The value of \SI{10}{\%} was roughly determined to give the simulation and experimental data equal weights when constructing the surrogate model. For the remainder of this work, we selected the model with $\sigma_\text{add} = \SI{10}{\percent}$ applied to the simulation data.

    One feature of the obtained surrogate model $\mathfrak{f}$, which is used for reconstructing the response function $c(E_\text{n})$, is that it is differentiable and modular. In this case, modular means that if more data becomes available, the model can be refitted easily to extend the data range and increase its predictive capabilities.

    Crucially, the model $\mathfrak{f}$ is trained on the logarithmic values of the supplied response data. This is done to achieve greater stability during the training, as the data points span almost six orders of magnitude. Thus, the model returns the logarithmic response function $c_\text{sm}$
    \begin{align}
        c_\text{sm}(E_\text{n}) = \mathfrak{f}\left( \log_{10}{(E_\text{n})}\right)
    \end{align}
    and the reconstructed model response function, as depicted in \autoref{fig:bootstrap} (red line), is obtained by
    \begin{align}
        c(E_\text{n}) &= 10^{c_\text{sm}(E_\text{n})} \nonumber \\
        &= 10^{\mathfrak{f}\left( \log_{10}{(E_\text{n})}\right)}.
        \label{eq:c_surrogate}
    \end{align}

    The applied model attributes data points with lower uncertainty a higher weight during the fitting, meaning the model prioritizes these values, which is the reason for the bump in $c(E_\text{n})$ in \autoref{fig:bootstrap} between approximately \SIrange[range-units=single]{400}{800}{keV}.
    
    The derivative, which is needed to determine the uncertainty of the calculated neutron number $N$ according to \autoref{eq:uncert-partial}, is given by
    
    \begin{align}
        \frac{\partial c}{\partial E_\text{n}}\left( E_\text{n}\right) =    \frac{\mathfrak{f}^\prime \left( \log_{10} (E_\text{n}) \right)}{E_\text{n}} \cdot 10^{\mathfrak{f} \left( \log_{10}(E_\text{n}) \right)}.
        \label{eq:c-eff-deriv}
    \end{align}

    The derivative of the fit function $\mathfrak{f}^\prime$ can be obtained directly from the surrogate model.
    
    \subsection{Determination of the effective \textit{c} \label{sec:c_eff}}
    To calculate $c_\text{eff}$, we convolute the output of the surrogate model for $c(E_\text{n})$, according to eq. \ref{eq:c_surrogate}, with the differential neutron flux, $\phi (E_\text{n}) = \di \varphi/\di E_\text{n}$. In this study, the neutron spectrum is obtained from Monte Carlo simulation; see \autoref{sec:results}. However, an experimentally determined spectral shape of the neutron beam can also be used as long as it is normalized. $c_\text{eff}$ can then be calculated by  
    
    \begin{align}
        c_\text{eff} = \frac{1}{\eta} \cdot \int_{E_0}^{E_u} c(E_\text{n}) \cdot \phi (E_\text{n}) \di E_\text{n}
        \label{eq:c_eff}
    \end{align}

    Where $E_0$ and $E_u$ are the upper and lower energy limits for which $c(E_\text{n})$ or $\phi (E_\text{n})$ is defined. $\eta$ is the normalization factor, given by the total conversion efficiency from incident source particle to neutrons, given in neutrons per incident source

    \begin{align}
        \eta = \int_{E_i}^{E_f} \phi (E_\text{n}) \di E_\text{n},
    \end{align}

    where $E_i$ and $E_f$ are the lowest and highest neutron energies recorded in the spectrum, respectively. 
    Normalizing the spectrum is important to obtain realistic values for $c_\text{eff}$. 
    It should be noted that the normalization factor does not need to be obtained from Monte Carlo simulations but can be obtained from experimentally measured spectra as well.
    In general, $E_i$ and $E_f$ do not need to equal $E_0$ or $E_u$. 
    In our case, the PHITS simulation gives the maximum and minimum energy of our energy interval, and, therefore, $E_0=E_i$ and $E_u=E_f$. 
    Often times this might not be the case, for example one exception is the case in which the energy range of $\phi (E)$ exceeds that of $c(E_\text{n})$. To highlight the influence of the spectral shape of the generated neutrons on the resulting value for $c_\text{eff}$, we determine $c_\text{eff}$ for four different conversion materials, commonly discussed in the context of LDNS in \autoref{sec:comparison}. 
    
    The variance of $c(E_\text{n})$ obtained from the surrogate model follows a Gaussian distribution around the model's value. The same applies to the variance of the neutron flux, as tallied by PHITS. We then calculate the total variance of $c_\text{eff}$ from the convolution of the two Gaussian distributed individual uncertainties. This leads to $\sigma_\text{total}^2 = \sum_E \sigma_{E,\text{Model}}^2 + \sigma_{E, \text{PHITS}}^2$, where $E$ stands for each energy bin. The relative uncertainty is then given by $\Delta c_\text{eff}/c_\text{eff} = \sqrt{\sigma_\text{total}^2}$.

    Furthermore, the described approach to determine $c_\text{eff}$ is also applicable when different projectile species are present. For example, at laser-driven neutron sources, it is common to use protons and deuterons to generate neutrons. In this case, the neutron spectrum used in \autoref{eq:c_eff} can be described as a linear combination of the relative contributions
    \begin{equation}
        \Phi_\text{tot} (E_\text{n}) = \sum_i \alpha_i \phi_i(E_\text{n}),
    \end{equation}
    with $\sum_i \alpha_i = 1$. $\alpha_i$ represent the relative contribution of projectile $i$ to the total yield, whereas $\phi_i(E_\text{n})$ is the spectral neutron distribution.

\section{Results and Discussion \label{sec:results}}
    \subsection{Comparison of the effective \textit{c} calculated by different methods \label{sec:comparison}}
\begin{figure}
        \centering
        \includegraphics[width=0.42\textwidth]{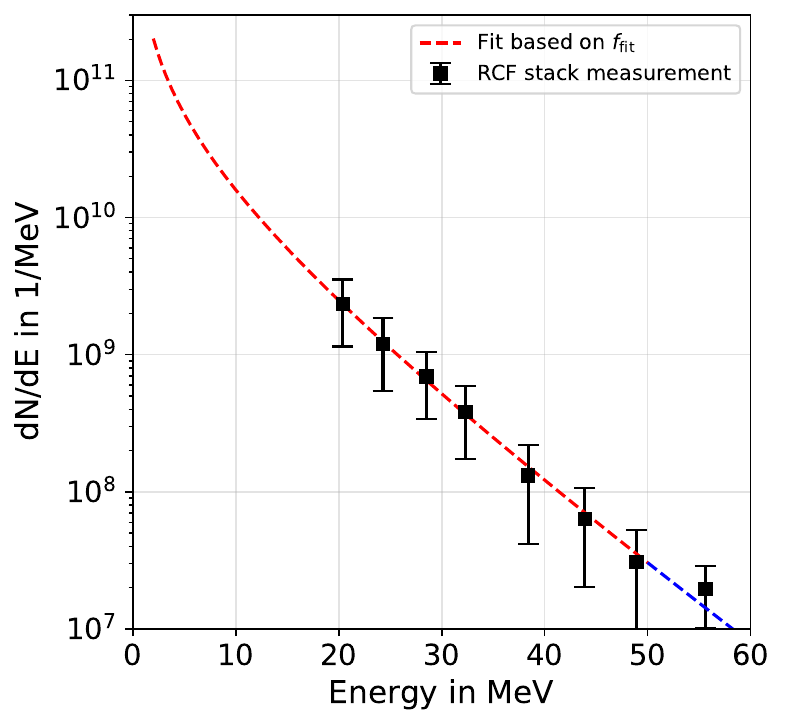}
        \caption{RCF measurement of the generated proton beam at DRACO, indicated by squares. Also shown is the fit based on \autoref{eq:proton-spectrum_fit}. The function is fitted to the black data points, indicated by the blue and red dashed lines. The spectrum used in the Monte Carlo simulations is truncated at \SI{49.8}{MeV} and represented by the red (dashed) line, as discussed in the text. All subsequent Monte Carlo simulations use this fit as the proton input spectrum.}
        \label{fig:proton_spectrum}
    \end{figure}

    The proposed method is evaluated using a proton spectrum measured during an experimental campaign at the DRACO PW laser system at the HZDR in Dresden, Germany. 
    Here, protons were accelerated from \SI{200}{nm} thin plastic foils in the TNSA regime. 
    The spectrum is obtained from RCF measurements and is extrapolated by fitting a function of the form 
    \begin{align}
        f_\text{fit}(E_\text{p}) = \frac{N_0}{E_\text{p}}\cdot \text{e}^{-E_\text{p}/T_\text{p}}
        \label{eq:proton-spectrum_fit}
    \end{align}
    to the data. 
    For more details on this method, the reader is conferred to the work of Schmitz \textit{et al.} \cite{Schmitz_2022} or Nuernberg \textit{et al.} \cite{Nuernberg_2009}.
    Here, $E_\text{p}$ is the kinetic energy of the protons, and $T_\text{p}$ is the temperature of the proton spectrum in MeV. 
    Both $N_0=\num{5.1 \pm 0.4 e11}$ and $T_\text{p} = \SI{8.6 \pm 0.3}{MeV}$ are obtained from the fit. 
    \autoref{fig:proton_spectrum} shows the fit and the measured data. 
    During the campaign an average proton cut-off energy of \SI{50 \pm 4}{MeV} was recorded, thus the fit is truncated at this energy, with a low energy limit of \SI{2}{MeV}. 
    The lower limit is based on the threshold energy for (p,n) reactions in commonly used converter materials at LDNS.
    
\begin{figure}
        \centering
        \includegraphics[width=0.42\textwidth]{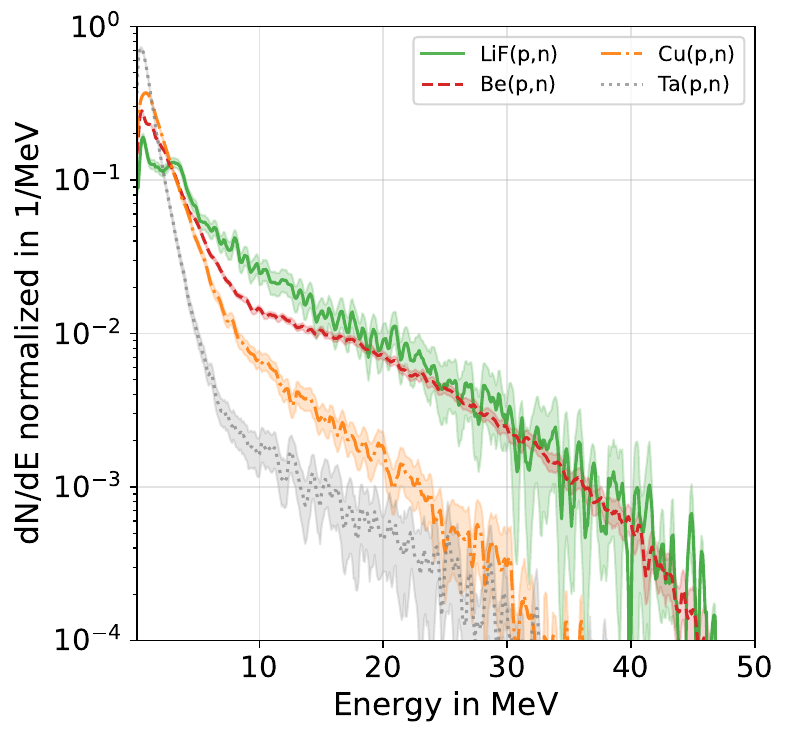}
        \caption{Normalized neutron spectra scored at \SI{0}{\degree} relative to the incoming proton beam for different catcher materials. The spectra are obtained from PHITS simulations and normalized by the total neutron production for each respective catcher. Shaded areas represent the statistical uncertainty. In this plot, LiF refers to the LiF catcher covered by a \SI{30}{\upmu m} steel foil.}
        \label{fig:neutron_spectra}
    \end{figure}

    The proton spectrum is used as an input for the PHITS simulations to calculate the generated neutron flux and afterwards $c_\text{eff}$ based on the approach discussed in \autoref{sec:Methods}. More details on the Monte Carlo simulations can be found in appendix \ref{app:phits-mc}. For neutron generation, different conversion targets are tested. Converters based on Lithium-Fluoride (LiF) and Beryllium (Be) have a cylindrical shape with a radius of \SI{1.9}{cm} and a thickness of \SI{1}{cm}. Additionally, \SI{3}{mm} thick plates of Copper (Cu) and Tantalum (Ta), with $\SI{5}{cm} \times \SI{5}{cm}$ dimensions, are also evaluated and compared to the rest. During the experimental campaign, only the LiF and Cu catcher were used, and we added the other two materials as a point of comparison. This catcher's geometric shape is chosen according to the catchers used during the experiment. 
    Furthermore, it should be noted that during the experiment, we placed a \SI{30}{\upmu m} thick steel foil in front of the LiF catcher to reduce debris generation. This foil was not necessary when using the Cu catcher. 
    We included the \SI{30}{\upmu m} steel foil in the Monte Carlo simulations for LiF.
    Neutron spectra in the simulation are recorded by a tally placed at \SI{0}{\degree} relative to the proton beam and can be seen in \autoref{fig:neutron_spectra}.
    Placing the steel foil in front of the LiF resulted in a $\sim \SI{13}{\%}$reduction in neutron emission at \SI{0}{\degree}, compared to a LiF catcher that is not covered by the steel foil, according to the Monte Carlo simulations. The shape of the emitted neutron spectrum at \SI{0}{\degree} is not significantly altered compared to the case without the steel foil. From this point, going further, results labelled as LiF refer to the combination of the steel foil placed in front of the LiF catcher unless stated otherwise.
    \autoref{tab:c_eff-comp} displays values for $c_\text{eff}$ using our revised method, based on the neutron spectra shown in \autoref{fig:neutron_spectra}.
    As a point of comparison, commonly used values/approaches taken from different publications \cite{Kleinschmidt2018, jung2013characterization} are also included. In Kleinschmidt \textit{et al.} \cite{Kleinschmidt2018}, the response $c_\text{eff}$ of bubble detectors is averaged from \SIrange[range-units=single]{0.3}{30}{MeV}, thus only one value is used for all materials (and angles). The method used in Jung \textit{et al.} \cite{jung2013characterization} is similar to this work and based on the convolution of the neutron spectrum with the response function. However, the interpolation of the response function is based on data published in Olsher \textit{et al.} \cite{olsher2007high} and Smecka \textit{et al.} \cite{smecka2007neutronendosimetrie}, which is a different, more limited data set compared to the one used in this work, "only" reaching up from \SIrange[range-units=single]{0.1}{200}{MeV}. The data points and interpolation are listed in appendix \ref{app:jung-response}.\\

    \begin{table}
        \centering
        \caption{Calculated values for $c_\text{eff}$ in $\times 10^{-5} ~\text{b} \cdot \text{cm}^2$/neutron for converter targets made from either Be, LiF, Cu or Ta. The materials are selected as examples based on their relevance to the field of LDNS. Neutron spectra used in the convolution are obtained from Monte Carlo simulations, where the spectrum is tallied under \SI{0}{\degree} relative to the incoming proton beam. The results from the method described in this article are compared to values obtained from other publications. As stated in the text, LiF refers to the combination of the LiF catcher covered by a \SI{30}{\upmu m} thick steel foil.}
        \begin{tabular}{c|c|c|c} 
            \hline \hline
            Material & This work & Kleinschmidt \textit{et al.} \cite{Kleinschmidt2018} & Jung \textit{et al.} \cite{jung2013characterization} \Tstrut\Bstrut \\ 
            \hline
            LiF & $4.13 \pm 0.39$ & $3.0 \pm 0.4$ & $3.30 \pm 0.11$ \Tstrut \\
            Be & $4.19 \pm 0.22$ & $3.0 \pm 0.4$ & $3.23 \pm 0.05$ \\
            Cu & $4.25 \pm 0.25$ & $3.0 \pm 0.4$ & $3.13 \pm 0.09$ \\ 
            Ta & $4.33 \pm 0.27$ & $3.0 \pm 0.4$ & $2.84 \pm 0.09$ \Bstrut  \\
            \hline \hline
        \end{tabular}
        \label{tab:c_eff-comp}
    \end{table}
    Starting by comparing the values of $c_\text{eff}$ displayed in \autoref{tab:c_eff-comp} for the method proposed in this work to the approach of Kleinschmidt \textit{et al.} reveals an increase of \SIrange[range-units=single]{38}{44}{\%}. The resulting difference in $c_\text{eff}$ mainly originates from the weighting of the response function by the neutron spectrum, with a less meaningful contribution arising from the energy range extending beyond the $\SI{0.3}{MeV} \leq E_\text{n} \leq \SI{30}{MeV}$ interval. Due to the neutron number per solid angle $N \propto 1/c_\text{eff}$, according to \autoref{eq:bubbles}, this results in an equal reduction in measured neutron flux. Based on the method of Kleinschmidt \textit{et al.}, the neutron flux using a Ta converter target would be \SI{44}{\%} higher compared to the method proposed herein. As can be seen, using an average value over a wide range of neutron energies leads to significant deviations, as the influence of differences in the emitted neutron spectra is lost. 
    Our proposed method results in a difference of up to \SI{5}{\%} in the value of $c_\text{eff}$ by changing the material of the catcher, which is within the uncertainty of the model. However, depending on the value chosen for $\sigma_\text{add}$ (discussed in \autoref{sec:cdetermination}), the difference between in $c_\text{eff}$ for the tested materials can reach up to $\sim \SI{20}{\%}$ (see \autoref{fig:delta_ceff}).

    Utilizing the approach of Jung \textit{et al.} also shows a dependency of $c_\text{eff}$ on the emitted spectral shape of the generated neutrons, recording a maximum deviation between the tested materials of up to \SI{16.2}{\%}. Details on the response function used by Jung \textit{et al.} can be found in \autoref{app:jung-response}. It should be noted that the uncertainty of $c_\text{eff}$ is much smaller compared to our approach. This is due to the absence of an uncertainty band for the interpolated response function shown in \cite{jung2013characterization}. Interestingly, the highest value for $c_\text{eff}$ is obtained for the LiF converter and the lowest is obtained using Ta. This is in contrast to our method, which returns the lowest values for $c_\text{eff}$ when using LiF and the highest for Ta. The difference can be explained when investigating the response function in Jung \textit{et al.} \cite[Figure 4]{jung2013characterization} more closely, where it can be seen that the interpolation fails to accurately reproduce the response of bubble detectors below \SI{300}{keV}. In this energy range, the interpolated response function significantly overestimates the actual response of the bubble detectors, leading to a much more significant contribution of low-energy neutrons. Since Ta generates the most low energy neutrons, see \autoref{fig:neutron_spectra}, this overestimation impacts it the most, followed by Cu and Be. Thus, neutron numbers calculated by this method would increase by up to \SI{52.4}{\%} if the catcher is made from Ta, for example.

    In \autoref{app:systematicuncertainty}, the influence of different values for $\sigma_\text{add}$ on the value for $c_\text{eff}$ is shown. Comparing the results obtained from the response function based on the unmodified simulation uncertainty data ($\sigma_\text{add} = 0$) to the modified data ($\sigma_\text{add} > 0$) results in a maximum deviation ranging from \SI{7}{\%} (Be) to \SI{20}{\%} (Ta). The deviation is a consequence of the differences between the measured and simulated response function from \SIrange[range-units=single]{0.4}{1}{MeV}. Since Ta and Cu have the largest emission in this energy range, see \autoref{fig:neutron_spectra}, these materials are most affected by this discrepancy. 

    To summarize, our method for taking the neutron-energy dependency of the response function into account results in differing values for $c_\text{eff}$ compared to previously established methods. Depending on the weight attributed to the simulations results versus the measurement results, the obtained values for $c_\text{eff}$ can vary by up to \SI{8}{\%} for the tested materials. Since the underlying response function data is almost 30 years old at this point, it should be updated with new measurements/simulations to improve the accuracy of our model and reconcile differences between the simulation and measurement results. This is especially important in the range from \SIrange[range-units=single]{0.4}{1}{MeV}, where a significant difference between simulated and measured response data can be seen.

    \subsection{Evaluation of neutron fluxes measured by bubble detectors \label{sec:exp-data}}
    To determine whether the method presented here or the referenced methods devised by Kleinschmidt \textit{et al.}\cite{Kleinschmidt2018} or Jung \textit{et al.}\cite{jung2013characterization} reproduce the integral neutron flux more faithfully, bubble detectors used at the DRACO laser are evaluated using each method and subsequently compared to in-depth Monte Carlo simulations. 
    The Monte Carlo simulations models the full experimental setup, including the room, target chamber, and shielding as depicted in the appendix in \autoref{fig:HZDRSetup}.
    
    All neutrons entering the tally at the bubble detector's position are scored, and their arrival time and energy are recorded. 
    In the experiment, the bubble detectors were placed at \SI{0}{\degree} relative to the TNSA ion beam and at a distance of \SI{77 \pm 1}{cm} to the catcher.
    Furthermore, the detectors were placed directly fixed to the target chamber walls. Two detectors were placed together, and the experimental neutron flux was determined by taking the average of the two detectors after they were irradiated for multiple shots to reduce the influence of shot-to-shot fluctuations. The irradiated bubble detectors were read using the BDR-III - Bubble Detector Reader manufactured by BTI \cite{bti2024reader}. Both Cu and LiF were used as conversion materials during the experimental campaign. In the case of Cu, data from \SI{100}{shots} are averaged, and LiF results are based on a total of \SI{50}{shots}.
    \begin{figure} 
        \centering
        \includegraphics[width=0.4\textwidth]{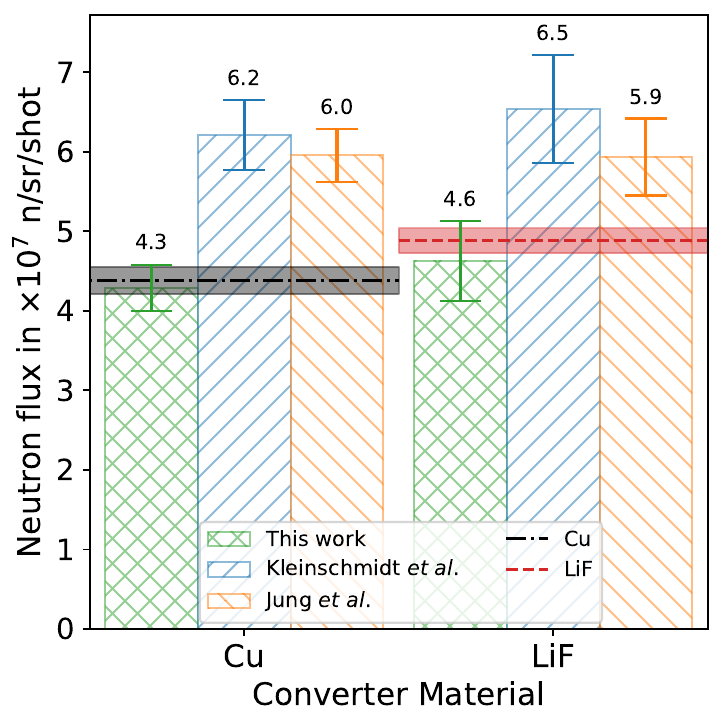}
        \caption{Comparison between the calculated neutron fluxes for a LiF and Cu catcher, based on the values shown in \autoref{tab:c_eff-comp} for $c_\text{eff}$. The bars represent the calculated neutron fluxes based on bubble detector measurements during the experiment. The detectors were placed at \SI{0}{\degree} relative to the proton beam. The horizontal lines indicate the neutron flux at the position of the detectors obtained from in-depth Monte Carlo simulations. The simulations included the entire geometry of the experimental setup and can be seen in \autoref{app:phits-mc}. The $2\sigma$ statistical uncertainty of the simulation is indicated by the coloured, shaded area.}
        \label{fig:methods_comp}
    \end{figure}   
    
    In \autoref{fig:methods_comp}, results using  LiF and Copper neutron production targets are shown. 
    Neutron fluxes calculated using the approach of this work are indicated by green bars. In contrast, the approach of Kleinschmidt \textit{et al.} is represented by blue bars, with results obtained by following Jung \textit{et al.} shown in orange. Also shown as a dashed line are the expected neutron fluxes from the Monte Carlo simulations of the experimental setup at the position of the bubble detectors.

    From the figure, it can be seen that the neutron flux is overestimated in comparison to the results obtained from the Monte Carlo simulations when employing the approach of either Kleinschmidt \textit{et al.} or Jung \textit{et al.}, due to the smaller value for $c_\text{eff}$ of the respective methods. This overestimation is more pronounced in the case of Cu, with an increase of \SI{46}{\%} and \SI{42}{\%} over the simulation results, respectively. This paper's revised approach to evaluate the experimental data results in a deviation of less than \SI{3}{\%}, and the integral neutron flux at the position of the detectors is accurately recovered. The same behavior is observed when switching to the LiF catcher. Here, the approach of Kleinschmidt \textit{et al.} overestimates the neutron flux by \SI{36}{\%} and the approach of Jung \textit{et al.} by \SI{23}{\%}. On the other hand, our approach underestimates the neutron flux by around \SI{5}{\%}. Nonetheless, the result of our approach falls within the uncertainty of the simulation results.

    Overall using our method results in lower neutron numbers by \SIrange[range-units=single]{22}{33}{\%} compared to previous methods due to higher values for $c_\text{eff}$. 
    Furthermore, the neutron fluxes calculated by our revised method more closely match results obtained from Monte Carlo simulations of the experimental setup. \\

    We conclude that the first test of our method looks promising with respect to determining neutron numbers per solid angle from bubble detector measurements. However, a more rigorous benchmarking campaign is needed in the future to verify these findings and highlight potential shortcomings of the proposed method.

    While our model's deviation from the simulation results is lower than the variants used for comparison, our method's precision (or uncertainty) must also be addressed. 
    We define this precision via the relative uncertainty $\Delta N/N$. $\Delta N$ is calculated according to \autoref{eq:uncert-partial}, with the full expression of the equation, including each individual uncertainty term given by \autoref{eq:delta-bubbles} in the appendix. Based on these equations, we estimate the intrinsic limitations of the precision of this work's approach. 

\begin{figure} [ht]
        \centering
        \includegraphics[width=0.4\textwidth]{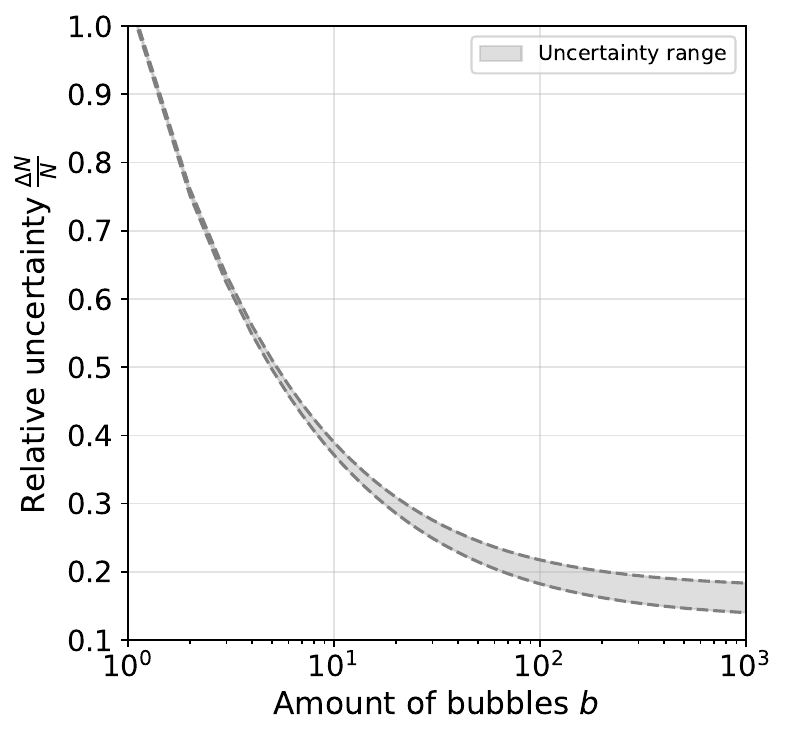}
        \caption{Precision of the neutron flux measurement for this work's approach in dependency on number of bubbles $b$. The grey area represents the uncertainty range for bubble detectors placed between 0.5 and \SI{3}{m} from the catcher.}
        \label{fig:uncert-b-d-inf}
    \end{figure}

    We investigate the precision in dependence on the number of bubbles $b$ inside a bubble detector for commonly used distances from the catcher to the detector. The distances we chose range from \SIrange[range-units=single]{0.5}{3}{m} and a sensitivity $s_0 = \SI{25}{bubbles/mrem}$ is assumed. Results are displayed in \autoref{fig:uncert-b-d-inf}. Concerning the number of bubbles, we set the maximum to \SI{1000}{bubbles}; values higher than this figure do not make sense from an experimental and readout point of view. The uncertainty in the number of bubbles $\Delta b = \sqrt{b}$ is based on Poisson statistics. Uncertainties for $\Delta c_\text{eff}$ and $\Delta s_\text{eff}$ are obtained from the procedures discussed in the main text and the appendix, and $\Delta E_\text{n}=\SI{0.05}{MeV}$ is determined by the energy grid used to tally the neutron spectrum in the simulations. Finally, $\Delta T=\SI{1}{\degree C}$ is set by the environment of the DRACO experimental area, and $\Delta d=\SI{3}{cm}$ is given by the setup and the geometry of the bubble detector itself.

    In \autoref{fig:uncert-b-d-inf}, it can be seen that the best-case relative uncertainty that can be achieved is around \SI{14}{\%} for a distance of \SI{3}{m} from the source and $b=\SI{1000}{bubbles}$. When positioning the detector at a distance of \SI{0.5}{m}, the lowest achievable uncertainty is around \SI{19}{\%}.
    Taking \autoref{eq:delta-bubbles} into account and looking at the results of \autoref{fig:uncert-b-d-inf}, it can be deduced that for $b \lesssim \SIrange[range-units=single]{20}{30}{bubbles}$, the relative uncertainty is mainly influenced by the Poisson uncertainty $\Delta b/b$ of the bubble number. In contrast, going to large values of $b \gtrsim \SI{200}{bubbles}$ shows a much slower decrease in the uncertainty, meaning that in this range, the dominant contribution does not originate from $\Delta b$. 
    Going as high is not supported by the bubble reader device. 
    The bubble detector reader is only certified for up to 150 bubbles\cite{bti2024reader}. 
    To achieve a better result evaluation capabilities have to be improved.
    Again, using \autoref{eq:delta-bubbles} shows that the contribution from $\Delta d$ is dominant in this range. Since $\Delta d$ is limited by the dimension of the bubble detector and the neutron source, this figure cannot be improved and thus represents the limit of the achievable precision. 

    These considerations establish a minimum relative uncertainty of $0.14 \leq \Delta N/N \leq 0.2$ under realistic experimental conditions. This figure is limited by the dimension of the detectors and neutron source and the number of bubbles that can reasonably count.
    
     
    \subsection{Influence of neutron scattering \label{sec:scatt}}
    As shown in the previous section, the energy dependence of the bubble detector response function can play a significant role when determining the neutron flux at the position of the detector. Under experimental conditions, neutrons that enter the detector do not necessarily take the shortest path from the catcher to the detector but can arrive at the detector via scattering. Depending on the geometry and materials of an experimental setup, neutron scattering can significantly influence the energy distribution of neutrons entering the detector. Furthermore, neutrons can be scattered into the detector that would have otherwise missed, resulting in an overestimation of the detected neutrons. Integral neutron measurements are especially sensitive to these scattered neutrons, as they cannot be distinguished from the overall signal.
    
\begin{figure}
        \centering
        \includegraphics[width=0.4\textwidth]{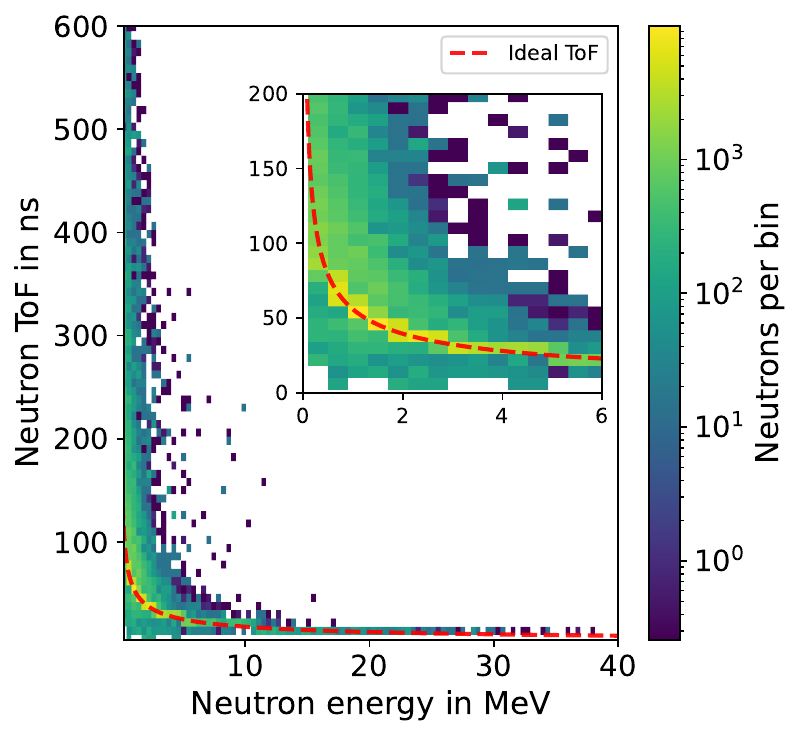}
        \caption{Neutron arrival time $t_\text{n}$ versus its energy $E_\text{n}$ at the position (\SI{0}{\degree}) of the bubble detector, when using the Cu conversion target. A significant amount of neutrons arrive via scattering, as indicated by contributions not coinciding with the ideal (no scattering events) relativistic time-of-flight, indicated by the dashed red line. The inset shows a zoomed-in view for $\SI{0}{ns} \leq t_\text{n} \leq \SI{200}{ns}$ and $\SI{0}{MeV} \leq E_\text{n} \leq \SI{6}{MeV}$.}
        \label{fig:Cu-RF_2D}
    \end{figure}

    In this section, we demonstrate the contribution of neutron scattering to our experimental setup by leveraging the setup's full geometry Monte Carlo simulations. In these simulations, neutrons that enter the detector unscattered are referred to as direct neutrons or the direct signal, whereas scattered neutrons are referred to as indirect neutrons/the indirect signal. To be able to distinguish between the two, the actual neutron time-of-flight (ToF) $t_\text{n}$ (arrival time at the detector) and the energy of the neutron $E_\text{n}$ are compared. \autoref{fig:Cu-RF_2D} shows the distribution of neutrons with respect to their arrival time and the energy with which they enter the detector volume. The dashed red line indicates the ideal/true time-of-flight for a given energy, meaning bins close to the line can be considered the direct signal. As can be seen, a significant number of neutrons arrive at times later than the true ToF.

    We can then distinguish between scattered and unscattered neutrons by using the correlation between energy and time-of-flight and calculating the time difference $\Delta t_\text{ToF}$, which we define as:

    \begin{equation}
        \Delta t_\text{ToF} = t_\text{n} - t(E_\text{n}).
    \end{equation}

\begin{figure} [ht]
        \centering
        \includegraphics[width=0.4\textwidth]{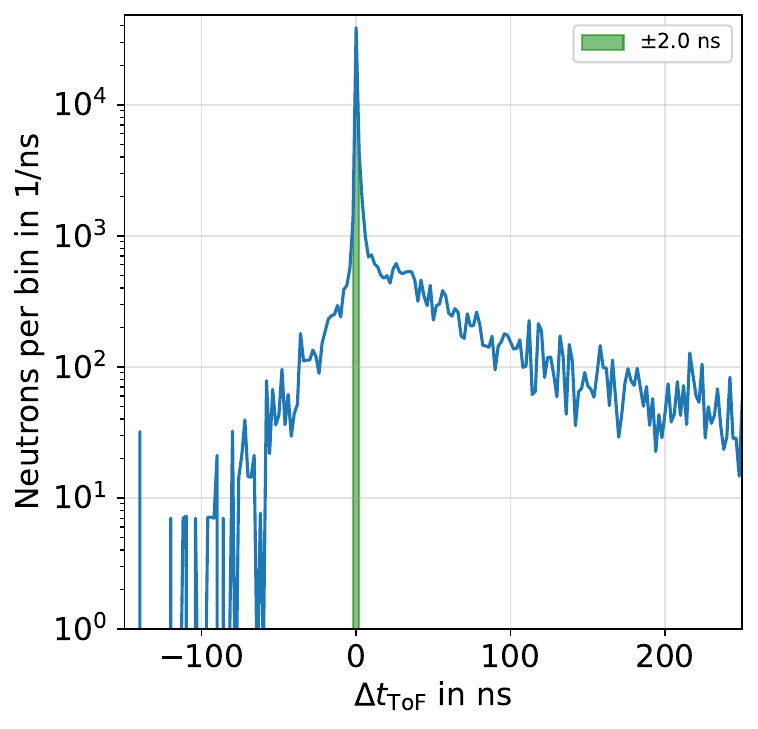}
        \caption{Neutron distribution of $\Delta t_\text{ToF}$ obtained from Monte Carlo simulations for the Cu conversion target. The shaded area indicates the time interval of $\di t = \SI{2}{ns}$, which we consider to contain the direct neutron signal, arriving unscattered.}
        \label{Cu-RF-1D}
    \end{figure}

    $t(E_\text{n})$ is the expected time that a neutron with energy $E_\text{n}$ needs to cover the distance from the source to the detector. $\Delta t_\text{ToF} = 0 \pm \di t$ means the neutron arrives at the detector at the expected ToF and is thus considered unscattered. In this context, $\di t$ is a time uncertainty due to the temporal pulse width of the source and path length differences. For our experimental setup, $\di t$ is around \SI{2}{ns} for the bubble detectors at \SI{0}{\degree}. \autoref{Cu-RF-1D} shows the neutron distribution in dependence on $\Delta t_\text{ToF}$. 
    
    Direct neutrons are contained within $\pm dt$, indicated by the green-shaded area. Taking the quotient of the shaded area's integral and the total integral of the distributions returns the percentage of direct neutrons arriving at the detector. Only \SI{53}{\%} of neutrons arrive unscattered for the Cu conversion target. When using the LiF conversion target, \SI{65}{\%} of neutrons arrive unscattered. These contributions are specific to each experimental setup and must be recalculated whenever the setup changes significantly or when measuring at a different facility. Thus, neutron fluxes measured at different facilities by bubble detectors only have limited significance if the scattering component is not properly accounted for.

    \begin{figure}
        \centering
        \includegraphics[width=0.4\textwidth]{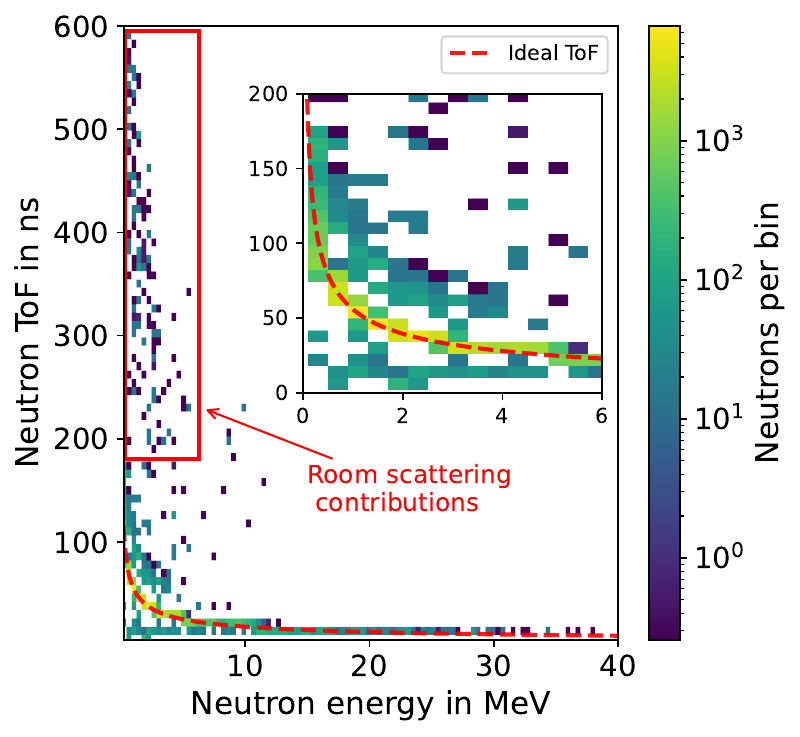}
        \caption{Neutron arrival time $t_\text{n}$ versus its energy $E_\text{n}$ at the position (\SI{0}{\degree}) of the bubble detector, when using the Cu conversion target. Here, neutrons that underwent at least one scattering event with the target chamber, its contents or the shielding are excluded, leading to a significant reduction of neutrons that deviate from the ideal ToF. With these neutrons excluded, \SI{90}{\%} of the scored neutrons fall within $\di t=\SI{2}{ns}$. Highlighted in the red box are neutrons that were only scattered by the room.}
        \label{fig:Cu-RF_2D_notc}
    \end{figure}

    Leveraging the counter function of the PHITS Monte Carlo code allows us to investigate the origin of the scattered neutrons. With counter functions, neutrons that underwent a scattering event inside a specific cell can be tagged, thereby allowing for differentiation between different scattering contributions after processing the simulation data. \autoref{fig:Cu-RF_2D_notc} shows the neutron energy versus arrival time when excluding neutrons that underwent at least one scattering event inside the target chamber (or its contents) and/or the shielding material of the experimental setup. From this figure, it could be determined that around \SI{90}{\%} of the scattering contribution originates from the target chamber (and its contents), as well as shielding. The contribution from neutrons scattering in the walls is less pronounced at around \SI{8}{\%}.
  
\section{Conclusion and outlook}\label{sec:conc}
    In this work, we propose and discuss a method for considering the neutron energy dependence of the bubble detector response function when converting measured bubble numbers to the corresponding neutron flux. Our presented model and the source code to convert bubble numbers to neutron fluxes are available for download and use on GitLab. The method is based on carefully reconstructing the bubble detector response functions by deploying a surrogate model, with subsequent convolution of the response function and the normalized neutron spectrum. The neutron spectrum can be obtained from experimental measurements or Monte Carlo simulations. For this work, we utilized simple Monte Carlo simulations, which only contained the conversion target and the proton beam, to obtain the neutron spectrum at the source in the direction of the bubble detector. \\
    
    Our method returns a reduced neutron flux compared to previously used methods, with a reduction of up to \SI{34}{\%} due to the weighted response function. Furthermore, we could demonstrate that neutron fluxes calculated from experimental data using our method are consistent with results obtained from Monte Carlo simulations of the full experimental setup. The method is tested for converter materials, LiF and Cu, deployed during the experiment. For considered cases, the deviation from the experimentally measured flux to the one obtained from the simulation was less than \SI{3}{\%} when using a copper conversion target and less than \SI{9}{\%} in the case of a LiF converter. We established a theoretical limit for our method's accuracy $\Delta N/N$ of about \SIrange[range-units=single]{15}{20}{\%}. From the accuracy analysis, we could determine that, for commonly used distances from the source to the detector, the number of bubbles per irradiation cycle should fall between \SIrange[range-units=single]{80}{200}{bubbles}. By accumulating bubbles in this range, the resulting relative uncertainty is approximately $\lesssim \SI{20}{\%}$, depending on the distance to the source. \\
    
    Finally, we highlighted the influence of neutron scattering on bubble detector measurements, an area that has not yet been thoroughly discussed in the context of laser-driven neutron sources. Scattered neutrons can contribute up to \SI{47}{\%} of the overall signal measured by bubble detectors. Due to the dependency of the scattered contribution on the experimental setup, comparisons between different laser facilities need to include scattering corrections. Neutrons can be scored multiple times depending on the setup if they scatter back and forth between the detector and its surroundings. During our measurement, the bubble detectors were fixed directly to the target chamber, commonly done at LDNS\cite{jung2013characterization, Kleinschmidt2018, zimmer2022demonstration}. Investigating the origin of the dominant scattering contributions for our experimental setup revealed that the target chamber and shielding, with minor components originating from room scattering. Therefore, we conclude that it is advisable to place bubble detectors away from the target chamber or bulk object, such as lead shielding, etc., to reduce scattering contributions. Since the contribution of room scattering was minor compared to the target chamber and surrounding materials, the best placement for the bubble detector is in a free-standing position. 
    
    Furthermore, since bubble detector are also susceptible to bubbles in the thermal/epithermal regime, it might be useful to shield the detectors with a proper shielding or collimator system for this energy range.

    This work is supposed to be a start of standardizing the usage of bubble detectors. 
    It is to be noted, that recurrent bubble detector sensitivity measurements would be great to increase the data set and verify the detectors capabilities.
    Furthermore the importance of supporting Monte Carlo simulations to characterize the neutron source and the potential scattering component is important.
    The developed method is, because of the differences between simulated and experimental data not ideal. Due to its data-driven nature it can be iteratively improved as soon as new data emerges, which over a longer period is supposed to increase the models quality.

\section*{Author's Contribution}
Conceptualization and simulations, S.S.; methodology, S.S. and B.S.; validation, S.S. and B.S.; writing—original draft preparation, S.S. and B.S.; measurement, analysis and discussion of experimental data: S.S., J.K., C.R., F.K., K.Z., M.A.M-C, A.A., C.G., T.J., A.J., B.S.; project lead, B.S.; funding acquisition, M.R.; All authors commented on the manuscript.
\begin{acknowledgments}
	This project is supported by the BMBF project "Zara-Lan" under grant number 15S9432 and by the EU project "IMPULSE" under grant number 871161 - PMOC-871161-7. It is also supported by the Graduate School CE within the Centre for Computational Engineering at Technische Universität Darmstadt. This project has received funding from the Euratom research and training program 2014-2018 under grant agreement No 847594 (H2020-Euratom-ARIEL), the Marie Curie actions of the 7th Framework Program under the grant agreement No 334315 (FP7-PEOPLE-NeutANdalus) and the Spanish national project PID2021-123879OB-C21. This work was supported by "la Caixa" Foundation (ID 100010434) (fellowship code LCF/BQ/PI20/11760027), Xunta de Galicia grant ED431F2023/21, and grant 
	RYC2021-032654-I funded by MICIU/AEI/10.13039/501100011033 and by "European Union NextGenerationEU".
	
	The authors thank the BTI employees and BTI itself for making their experimental and simulation data available to us. We thank Tatsuhiko Ogawa and the PHITS team for helpful discussions regarding the accuracy of the neutron generation simulations.
	Stefan Scheuren acknowledges support from Trumpf GmbH \& Co. KG.
\end{acknowledgments}

\section*{Data Availability Statement}
The data that support the findings of this study are openly available in TUdatalib at
\url{https://tudatalib.ulb.tu-darmstadt.de/handle/tudatalib/4259}. 
The current developments are listed on \url{https://git.rwth-aachen.de/surrogat-models/bubbledetectors}

\section*{Competing interest}
The authors declare no competing interests.

\appendix

   \section{Determination of the effective sensitivity \label{app:s_eff}}
    Besides the energy dependency discussed in the main text, the sensitivity of bubble detectors, i.e. the amount of bubbles generated per dose, depends on the temperature $s_0(T)$. This dependency of the sensitivity was investigated and has been published by BTI \cite[Figure 4]{Ing1997}\cite{Buckner1994}, with a reference temperature of \SI{20}{\degree C}. In the publications, the influence of the temperature is expressed by a multiplicative factor, which is why we define the temperature dependency as follows

    \begin{align}
        s_0(T) = s_0 \cdot f(T).
        \label{eq:temps-sens}
    \end{align}
    
\begin{figure}
        \centering
        \includegraphics[width=0.49\textwidth]{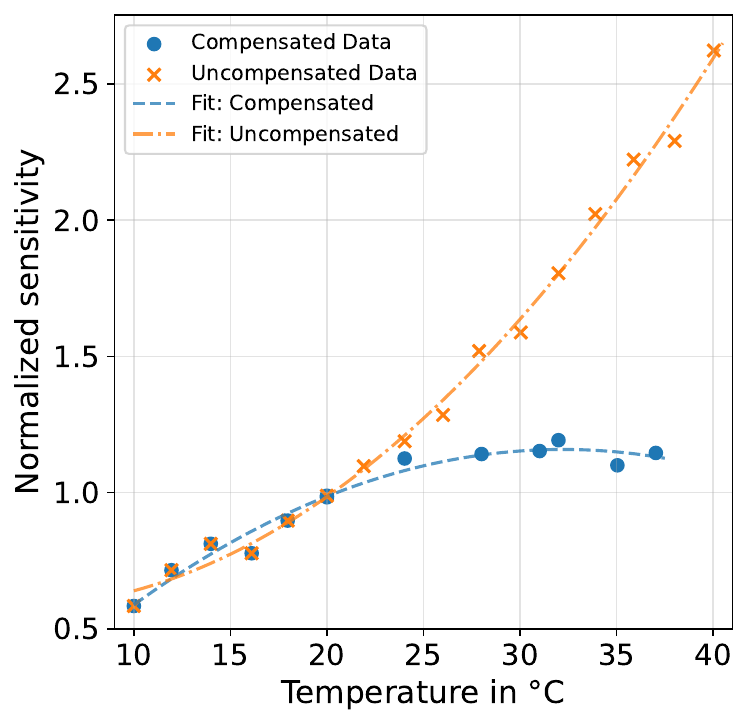}
        \caption{Temperature dependency for the two types of bubble detectors manufactured by BTI. Compensated/Uncompensated refers to the temperature dependency, where the compensated type tries to minimize the influence of the temperature on the sensitivity $s_0$ of the detectors. The dashed and dotted lines fit the data points based on \autoref{eq:temp-fit}. Data obtained from publications by BTI \cite{Buckner1994, Ing1997}.}
        \label{fig:temperatureDependence}
    \end{figure}

    We use the published data to fit a polynomial function $f(T)$ to the data to calculate the correction factor in \autoref{eq:temps-sens}. \autoref{fig:temperatureDependence} shows the temperature dependency for two different types of bubble detectors. One compensates for the temperature dependency as well as possible, and one is uncompensated. The data for both types are fitted using \autoref{eq:temp-fit}.
    
    \begin{align}
        f(T) &=\sum_{k=0}^n c_k\cdot T^k 
        \intertext{with $n=2$:}
             &= c_0 + c_1\cdot T + c_2\cdot T^2
        \label{eq:temp-fit}
    \end{align}
    
    For both data sets, the parameters can be determined by least squares fit as given in \autoref{tab:fitTemperature}.
    
    \begin{table}[ht]
        \centering
        \caption{Temperature coefficients from the temperature fit for the compensated and the uncompensated case.}
        \label{tab:fitTemperature}
        \begin{tabular}{l|rrr}\hline\hline
             & $c_0$ & $c_1$ & $c_2$ \\[1ex]
            Units  & 1 & 1/\si{\celsius} & 1/ \si{\celsius\squared} \\ \hline\\[.5ex]
            Comp   & -0.0411 & 0.0744 & -0.0012 \\
            Uncomp & 0.6004 & -0.0113 &  0.0015 \\\hline\hline
        \end{tabular}
    \end{table}

    Measurements conducted with bubble detectors in the context of this work use the compensated type, and the temperature sensitivity is thus accounted for by the fit to the compensated data points in \autoref{fig:temperatureDependence}, indicated by the green dashed line. \\
    
\begin{figure}[ht]
        \centering
        \includegraphics[width=0.4\textwidth]{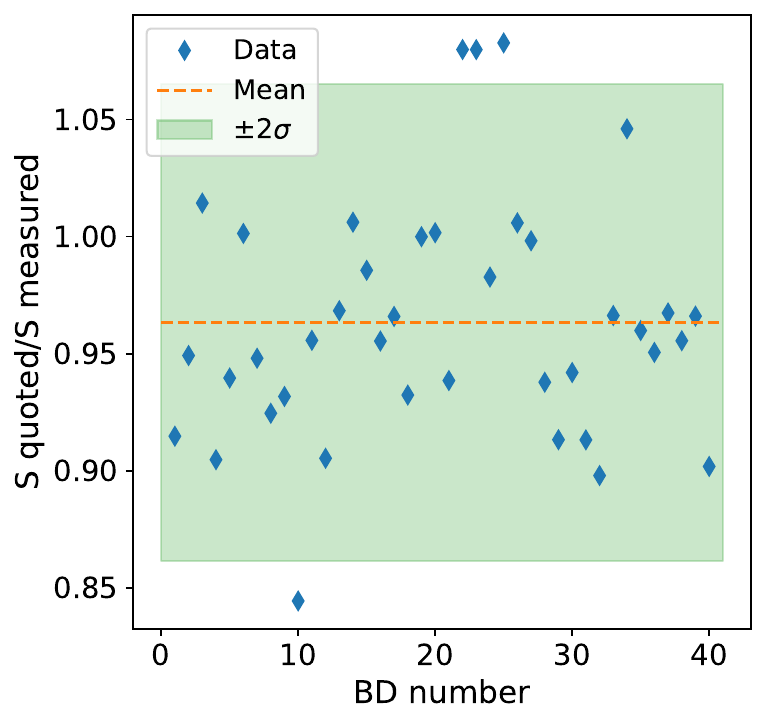}
        \caption{Comparison of the quoted and measured sensitivity for 40 different bubble detectors. Data obtained from Ing \textit{et al.} \cite[Figure 3]{Ing2001}.}
        \label{fig:sens-var}
    \end{figure}

    Lastly, bubble detectors, or rather the sensitivity, are subject to batch deviations, which can result in deviations from the quoted (supplied by BTI) to the actual sensitivity. This can be overcome by conducting a calibration measurement with a characterized neutron source. In a previous publication \cite[Figure 3]{Ing2001} BTI investigated deviation from the quoted to the measured sensitivity, the results of which can be seen in \autoref{fig:sens-var}.
    
    To account for these deviations, we apply a correction term $k_\text{dev}$ to the quoted sensitivity value $s_0$, i.e. $\hat{s}_0 = s_0/k_\text{dev}$. we  define the factor $k_\text{dev}$ as the mean of the data shown in \autoref{fig:sens-var}, which is indicated by the orange dashed line and returns $k_\text{dev} = 0.963$. Furthermore, we define the uncertainty of the correction factor by the standard deviation of the data set published by BTI. We chose $2\sigma$, represented by the green shaded area in \autoref{fig:sens-var}, as the uncertainty to stay consistent with the definition of the uncertainty for $c(E_\text{n})$ displayed in \autoref{fig:bootstrap} (grey area). Using these definitions we arrive at $\Delta \hat{s}_0/\hat{s}_0 = 0.102$. 

    Using the temperature dependency $f(T)$, as well as the correction term $k_\text{dev}$ to account for batch variation, we define the effective sensitivity $s_\text{eff}$ in the following way:
    
    \begin{align}
        s_\text{eff}(T) &= \frac{s_0}{k_\text{dev}} \cdot f(T)\\
        &=\hat{s}_0\cdot f(T),
    \end{align}

    where, as a reminder, $s_0$ is supplied by BTI when obtaining the bubble detectors. If a calibration measurement is conducted before deployment of the detectors, the sensitivity obtained from the calibration can be used instead of $\hat{s}_0$.
    
    In the code we provided, temperature and sensitivity correction can be manually turned on or off independently. A deviating individual (or own) correction function/data can also be provided if desired.

    \section{Uncertainty calculation \label{app:uncert}}
    Here, the full expression for calculating the uncertainty of the neutron number according to \autoref{eq:uncert-partial} is given

    \begin{align}
            &\left(\Delta N\right)^2 = \sum_i \left(\frac{\partial N}{\partial x_i}\cdot \Delta x_i\right)^2 \notag \\
            &= \left( \frac{d^2\cdot \Delta b}{c_\text{eff}\cdot s_\text{eff}}\right)^2 + \left( \frac{2b\cdot d \cdot \Delta d}{c_\text{eff}\cdot s_\text{eff}}\right)^2 + \left( -\frac{b\cdot d^2\cdot \Delta c_\text{eff}}{c_\text{eff}^2\cdot s_\text{eff}}\right)^2 \notag \\
            &+ \left( -\frac{b\cdot d^2\cdot \Delta \hat{s}_0}{c_\text{eff}\cdot \hat{s}_0^2\cdot f(T)}\right)^2 + \left( - \frac{b\cdot d^2\cdot \Delta T}{c_\text{eff}\cdot \hat{s}_0 \cdot f^2(T)} \cdot \frac{\partial f(T)}{\partial T} \right)^2 \notag \\ 
            &+ \left( -\frac{b\cdot d^2 \cdot \Delta E_n}{c_\text{eff}^2 \cdot s_\text{eff}}\cdot \frac{\partial c_\text{eff}(E_n)}{\partial E_n} \right)^2. 
        \label{eq:delta-bubbles}
    \end{align}

    Values for $\Delta d$,$\Delta T$ and $\Delta b$ are determined from experimental measurements, whereas $\Delta E_\text{n}$ is defined by the energy mesh of the Monte Carlo simulations and $\Delta c_\text{eff}$ and $\Delta \hat{s}_0$ is derived according to the procedure described section \ref{sec:c_eff} and appendix \ref{app:s_eff}. Note, $\hat{s}_0$ and $\Delta \hat{s}_0$ can also be determined experimentally by conducting a calibration measurement before using the bubble detectors.

    \section{Monte Carlo setup \label{app:phits-mc}}
    We used the PHITS Monte Carlo code in version 3.31 for the results presented in this paper. Proton and neutron reactions are simulated using the JENDL-4.0 nuclear data library, while TENDL-2019 is used for photo-nuclear reactions. For the simulation of heavy ion-induced reactions, such as carbon-induced neutron production, we rely on JQDM-2.0, with an adjusted low energy cut-off threshold of \SI{2}{MeV/u}. Photon and electron transport are handled by the electron gamma shower module version 5 (EGS5), which comes pre-packaged with PHITS.

    \begin{figure}[ht]
        \centering
        \includegraphics[width=0.4\textwidth]{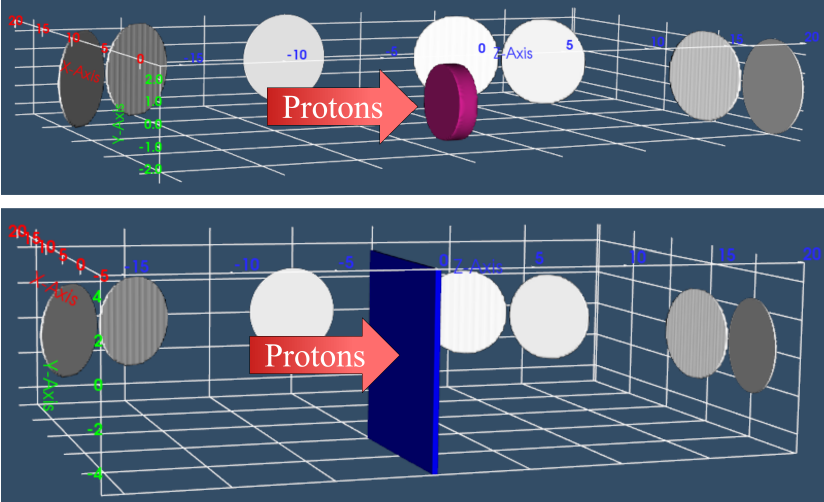}
        \caption{3D view of the simulation setup used to score the neutron spectra displayed in \autoref{fig:neutron_spectra}. The top shows the setup for the LiF and Be catcher, whereas the bottom shows the one using the Cu and Ta catcher. These spectra are then used to calculate $c_\text{eff}$. The \SI{30}{\upmu m} steel foil in front of the LiF catcher is not included in these simulations.}
        \label{fig:phits-setup-simple}
    \end{figure}    

    We conduct two sets of Monte Carlo simulations using PHITS for this study. The first set concerns generating neutron spectra from the incoming proton beam, shown in \autoref{fig:proton_spectrum}. These simulations only contain the conversion target made from LiF, Be, Cu and Ta and the neutron scorer used to obtain the spectrum. The tallies are placed at a distance of \SI{20}{cm} from the centre of the conversion target and have a circular cross-section with a radius of \SI{2.6}{cm}. In total, we place 13 detectors at different angles relative to the proton beam, ranging from \SIrange[range-units=single]{0}{180}{\degree} in increments of \SI{15}{\degree}. A 3D view of the simulation setup can be seen in \autoref{fig:phits-setup-simple}. Each simulation run used a total of \SI{5e10}{protons} to generate the neutron spectra.

    \begin{figure}[ht]
        \centering
        \includegraphics[width=0.4\textwidth]{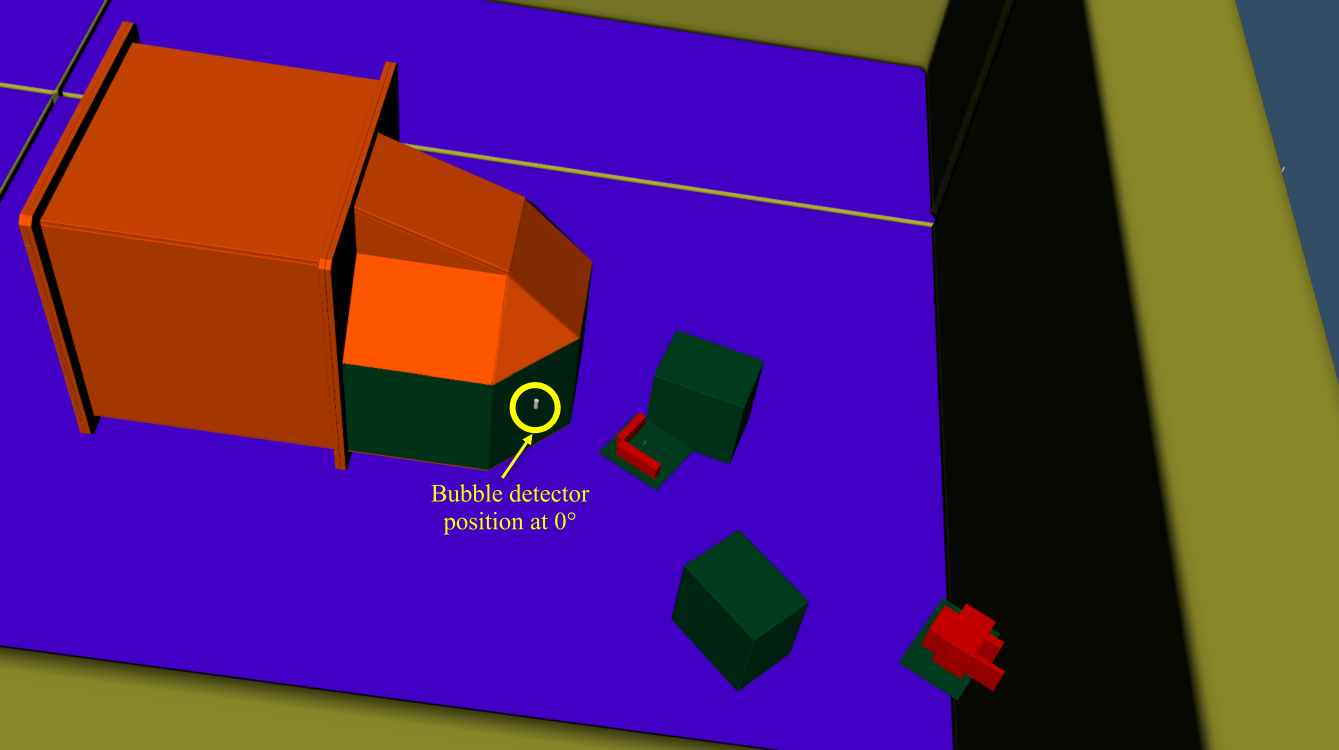}
        \caption{Experimental setup as modelled in PHITS. The bubble detectors were fixed to the target chamber walls within the yellow circle. Neutrons are generated inside the target chamber. Different colours correspond to different materials. Green is aluminium, orange represents stainless steel, red is used for lead, and beige is used for the concrete walls. Simulations using the geometry depicted here include the steel foil in front of the LiF catcher.}
        \label{fig:mc-full-geo}
    \end{figure}

    We refer to the second set of simulations as in-depth simulations. In these simulations, we modelled the most relevant parts of the experimental setup, including the room, the target chamber and the shielding setup. The setup can be seen in \autoref{fig:mc-full-geo}. We included neutron production from all relevant source terms associated with laser-based proton acceleration, such as neutrons generated by carbon ions, electrons or the counter-propagating proton beam. 
    
    \section{Response function used by Jung \textit{et al.} \label{app:jung-response}}
\begin{figure}[ht]
        \centering
        \includegraphics[width=0.4\textwidth]{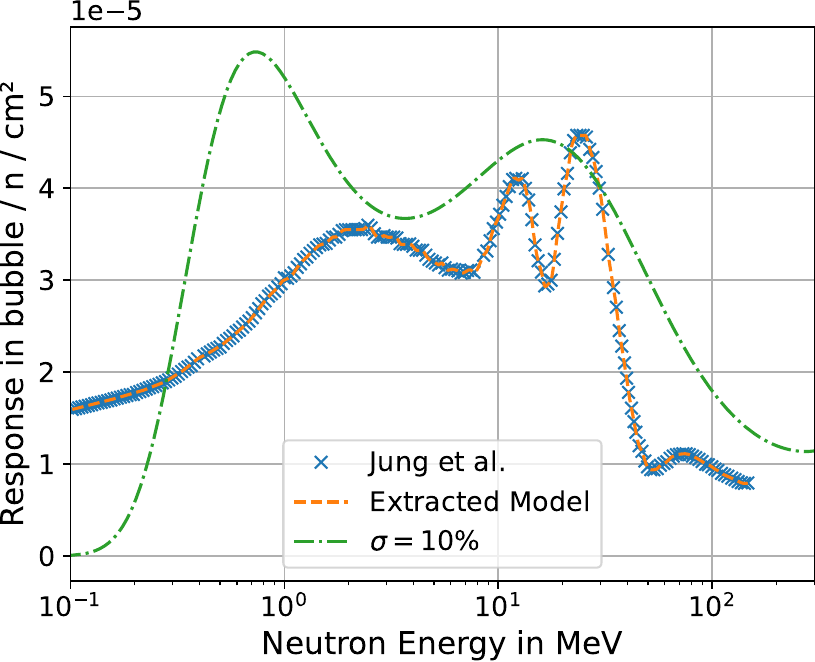}
        \caption{Interpolated response function data (blue markers) of Jung \textit{et al.} \cite{jung2013characterization} and cubic spline model fitted to the data (orange, dashed line). 
        The dotted green line indicates our surrogate with a systematic uncertainty for the simulation of \SI{20}{\percent}.}
        \label{fig:jung-model}
    \end{figure}

    Values calculated by the Jung \textit{et al.} method are based on the interpolated response function published in Jung \textit{et al.} \cite[Figure 4]{jung2013characterization}. We extracted the response function data from the plot and fitted it with a cubic spline function to obtain a value at arbitrary energies. The extracted data and the cubic spline fit can be seen in \autoref{fig:jung-model}. We do not include an uncertainty band, as no uncertainty was given in the original publication. 

    \section{Surrogate model with variations in the systematic uncertainty\label{app:systematicuncertainty}}
\begin{figure}
        \centering
        \includegraphics[width=0.4\textwidth]{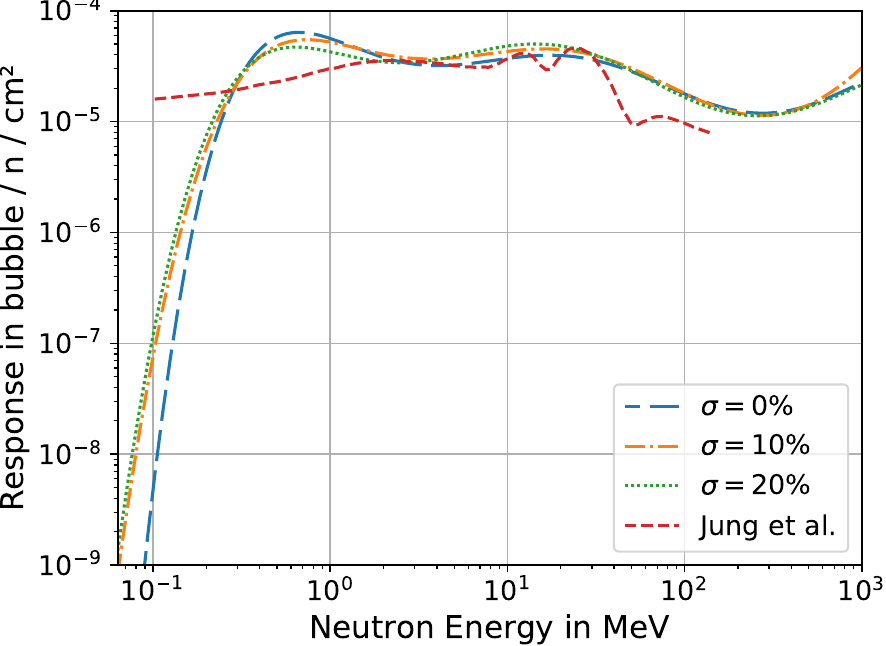}
        \caption{Comparison of the surrogate models with different systematic uncertainty factors and the model by Jung.}
        \label{fig:modelcompare}
    \end{figure}

    We investigated the influence of an additional uncertainty term resulting from the systematic uncertainty of the simulations.
    This was done by increasing the error before the resampling (as mentioned in \autoref{sec:cdetermination}) by a flat value. 
    Resulting models for a selection of different $\sigma_\text{add}$ are presented in \autoref{fig:modelcompare}.
    
\begin{figure}
        \centering
        \includegraphics[width=0.35\textwidth]{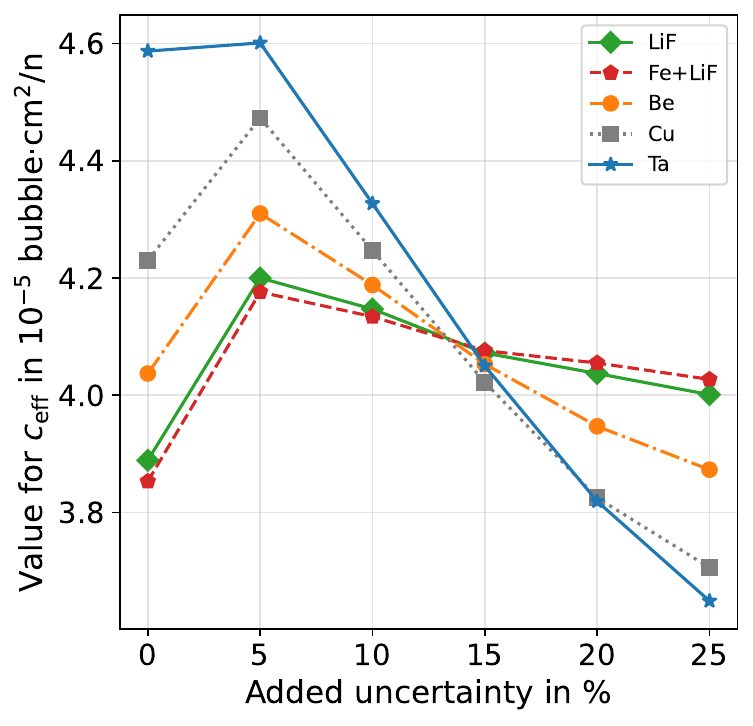}
        \caption{Calculated values for $c_\text{eff}$ for different values of $\sigma_\text{add}$. The lines connecting the markers are meant to guide the eye. Error bars are omitted for better readability.}
        \label{fig:delta_ceff}
    \end{figure}

    To better understand the influence of $\sigma_\text{add}$, we calculate the resulting $c_\text{eff}$ based on the different models, with results are presented in \autoref{fig:delta_ceff}. Going from smaller to larger $\sigma_\text{add}$ increases the influence that the experimentally measured response function has on the surrogate model.
    
    The magnitude of the maximum deviation between the unmodified $\sigma_\text{add} = 0$ and modified uncertainty data depends strongly on the materials of the converter. Be exhibits the lowest deviation at around \SI{7}{\%}, closely followed by LiF with \SI{8}{\%}. Cu and Ta, on the other hand, show large deviations exceeding \SI{10}{\%}, at 12 and \SI{20}{\%}, respectively. The largest deviations are observed for Cu and Ta due to the energy distribution of the emitted neutron spectrum, where more neutrons are emitted in the energy range most affected by different values for $\sigma_\text{add}$.
    \section{Experimental Setup at HZDR}
    A paper, discussing the full setup, with all displayed diagnostics is in preparation.
    \begin{figure} 
        \centering
        \includegraphics[width=0.5\textwidth]{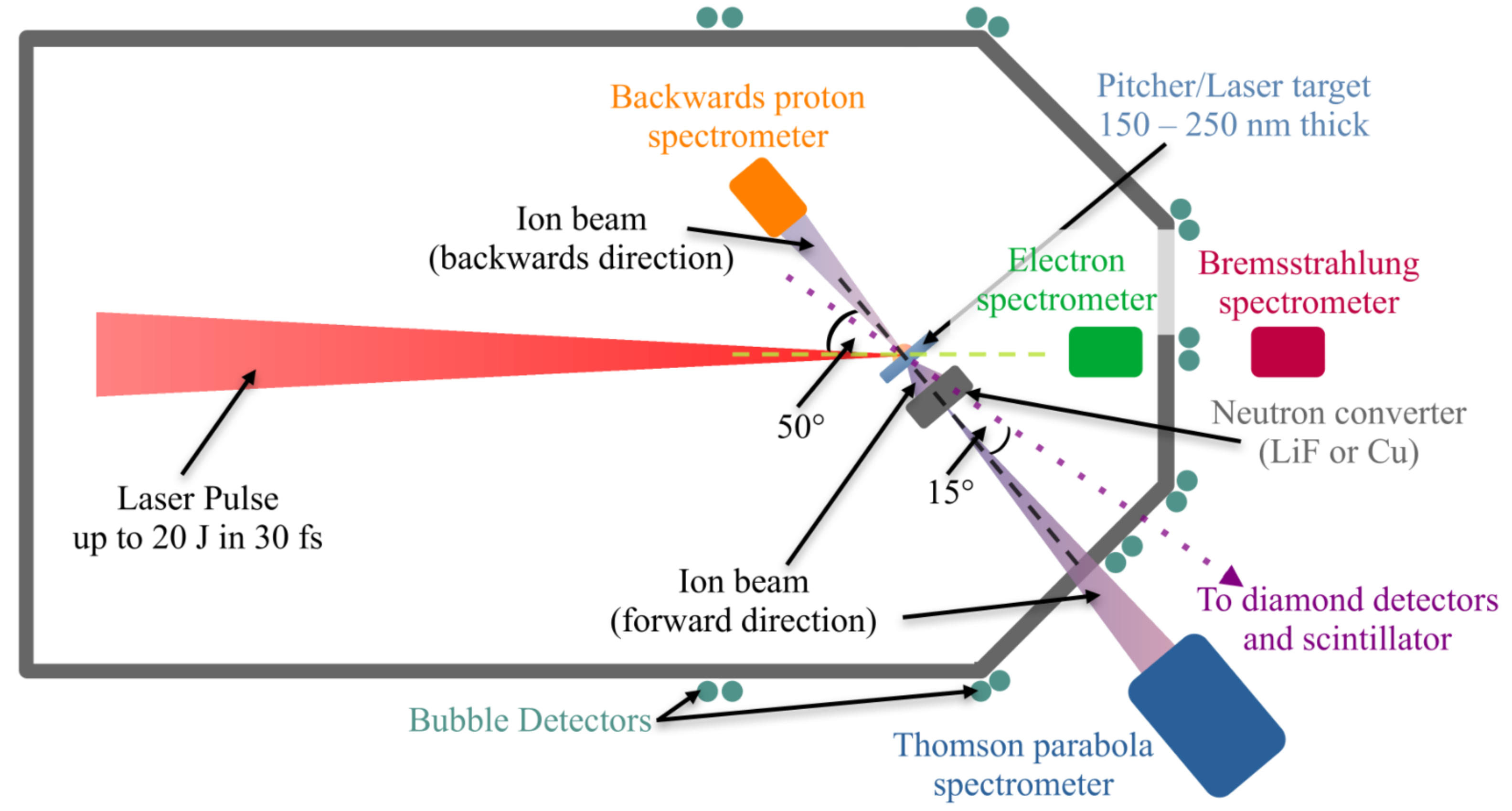}
        \caption{Experimental setup of the beamtime at HZDR. The bubble positions are indicated by the blue dots.}
        \label{fig:HZDRSetup}
    \end{figure}

\newpage
\bibliography{BubblePaper}

\begin{thebibliography}{37}%
\makeatletter
\providecommand \@ifxundefined [1]{%
 \@ifx{#1\undefined}
}%
\providecommand \@ifnum [1]{%
 \ifnum #1\expandafter \@firstoftwo
 \else \expandafter \@secondoftwo
 \fi
}%
\providecommand \@ifx [1]{%
 \ifx #1\expandafter \@firstoftwo
 \else \expandafter \@secondoftwo
 \fi
}%
\providecommand \natexlab [1]{#1}%
\providecommand \enquote  [1]{``#1''}%
\providecommand \bibnamefont  [1]{#1}%
\providecommand \bibfnamefont [1]{#1}%
\providecommand \citenamefont [1]{#1}%
\providecommand \href@noop [0]{\@secondoftwo}%
\providecommand \href [0]{\begingroup \@sanitize@url \@href}%
\providecommand \@href[1]{\@@startlink{#1}\@@href}%
\providecommand \@@href[1]{\endgroup#1\@@endlink}%
\providecommand \@sanitize@url [0]{\catcode `\\12\catcode `\$12\catcode
  `\&12\catcode `\#12\catcode `\^12\catcode `\_12\catcode `\%12\relax}%
\providecommand \@@startlink[1]{}%
\providecommand \@@endlink[0]{}%
\providecommand \url  [0]{\begingroup\@sanitize@url \@url }%
\providecommand \@url [1]{\endgroup\@href {#1}{\urlprefix }}%
\providecommand \urlprefix  [0]{URL }%
\providecommand \Eprint [0]{\href }%
\providecommand \doibase [0]{http://dx.doi.org/}%
\providecommand \selectlanguage [0]{\@gobble}%
\providecommand \bibinfo  [0]{\@secondoftwo}%
\providecommand \bibfield  [0]{\@secondoftwo}%
\providecommand \translation [1]{[#1]}%
\providecommand \BibitemOpen [0]{}%
\providecommand \bibitemStop [0]{}%
\providecommand \bibitemNoStop [0]{.\EOS\space}%
\providecommand \EOS [0]{\spacefactor3000\relax}%
\providecommand \BibitemShut  [1]{\csname bibitem#1\endcsname}%
\let\auto@bib@innerbib\@empty
\bibitem [{\citenamefont {Alvarez}\ \emph {et~al.}(2014)\citenamefont
  {Alvarez}, \citenamefont {Fern{\'a}ndez-Tobias}, \citenamefont {Mima},
  \citenamefont {Nakai}, \citenamefont {Kar}, \citenamefont {Kato},\ and\
  \citenamefont {Perlado}}]{alvarez2014laser}%
  \BibitemOpen
  \bibfield  {author} {\bibinfo {author} {\bibfnamefont {J.}~\bibnamefont
  {Alvarez}}, \bibinfo {author} {\bibfnamefont {J.}~\bibnamefont
  {Fern{\'a}ndez-Tobias}}, \bibinfo {author} {\bibfnamefont {K.}~\bibnamefont
  {Mima}}, \bibinfo {author} {\bibfnamefont {S.}~\bibnamefont {Nakai}},
  \bibinfo {author} {\bibfnamefont {S.}~\bibnamefont {Kar}}, \bibinfo {author}
  {\bibfnamefont {Y.}~\bibnamefont {Kato}}, \ and\ \bibinfo {author}
  {\bibfnamefont {J.}~\bibnamefont {Perlado}},\ }\href@noop {} {\bibfield
  {journal} {\bibinfo  {journal} {Physics Procedia}\ }\textbf {\bibinfo
  {volume} {60}},\ \bibinfo {pages} {29} (\bibinfo {year} {2014})}\BibitemShut
  {NoStop}%
\bibitem [{\citenamefont {Alejo}\ \emph {et~al.}(2016)\citenamefont {Alejo},
  \citenamefont {Ahmed}, \citenamefont {Green}, \citenamefont {Mirfayzi},
  \citenamefont {Borghesi},\ and\ \citenamefont {Kar}}]{alejo2016recent}%
  \BibitemOpen
  \bibfield  {author} {\bibinfo {author} {\bibfnamefont {A.}~\bibnamefont
  {Alejo}}, \bibinfo {author} {\bibfnamefont {H.}~\bibnamefont {Ahmed}},
  \bibinfo {author} {\bibfnamefont {A.}~\bibnamefont {Green}}, \bibinfo
  {author} {\bibfnamefont {S.}~\bibnamefont {Mirfayzi}}, \bibinfo {author}
  {\bibfnamefont {M.}~\bibnamefont {Borghesi}}, \ and\ \bibinfo {author}
  {\bibfnamefont {S.}~\bibnamefont {Kar}},\ }\href@noop {} {\bibfield
  {journal} {\bibinfo  {journal} {Nuovo Cimento C}\ }\textbf {\bibinfo {volume}
  {38}} (\bibinfo {year} {2016})}\BibitemShut {NoStop}%
\bibitem [{\citenamefont {Yogo}\ \emph
  {et~al.}(2023{\natexlab{a}})\citenamefont {Yogo}, \citenamefont {Arikawa},
  \citenamefont {Abe}, \citenamefont {Mirfayzi}, \citenamefont {Hayakawa},
  \citenamefont {Mima},\ and\ \citenamefont {Kodama}}]{yogo2023advances}%
  \BibitemOpen
  \bibfield  {author} {\bibinfo {author} {\bibfnamefont {A.}~\bibnamefont
  {Yogo}}, \bibinfo {author} {\bibfnamefont {Y.}~\bibnamefont {Arikawa}},
  \bibinfo {author} {\bibfnamefont {Y.}~\bibnamefont {Abe}}, \bibinfo {author}
  {\bibfnamefont {S.}~\bibnamefont {Mirfayzi}}, \bibinfo {author}
  {\bibfnamefont {T.}~\bibnamefont {Hayakawa}}, \bibinfo {author}
  {\bibfnamefont {K.}~\bibnamefont {Mima}}, \ and\ \bibinfo {author}
  {\bibfnamefont {R.}~\bibnamefont {Kodama}},\ }\href@noop {} {\bibfield
  {journal} {\bibinfo  {journal} {The European Physical Journal A}\ }\textbf
  {\bibinfo {volume} {59}},\ \bibinfo {pages} {191} (\bibinfo {year}
  {2023}{\natexlab{a}})}\BibitemShut {NoStop}%
\bibitem [{\citenamefont {Zimmer}\ \emph {et~al.}(2022)\citenamefont {Zimmer},
  \citenamefont {Scheuren}, \citenamefont {Kleinschmidt}, \citenamefont
  {Mitura}, \citenamefont {Tebartz}, \citenamefont {Schaumann}, \citenamefont
  {Abel}, \citenamefont {Ebert}, \citenamefont {Hesse}, \citenamefont
  {Z{\"a}hter} \emph {et~al.}}]{zimmer2022demonstration}%
  \BibitemOpen
  \bibfield  {author} {\bibinfo {author} {\bibfnamefont {M.}~\bibnamefont
  {Zimmer}}, \bibinfo {author} {\bibfnamefont {S.}~\bibnamefont {Scheuren}},
  \bibinfo {author} {\bibfnamefont {A.}~\bibnamefont {Kleinschmidt}}, \bibinfo
  {author} {\bibfnamefont {N.}~\bibnamefont {Mitura}}, \bibinfo {author}
  {\bibfnamefont {A.}~\bibnamefont {Tebartz}}, \bibinfo {author} {\bibfnamefont
  {G.}~\bibnamefont {Schaumann}}, \bibinfo {author} {\bibfnamefont
  {T.}~\bibnamefont {Abel}}, \bibinfo {author} {\bibfnamefont {T.}~\bibnamefont
  {Ebert}}, \bibinfo {author} {\bibfnamefont {M.}~\bibnamefont {Hesse}},
  \bibinfo {author} {\bibfnamefont {{\c{S}}.}~\bibnamefont {Z{\"a}hter}},
  \emph {et~al.},\ }\href@noop {} {\bibfield  {journal} {\bibinfo  {journal}
  {Nature communications}\ }\textbf {\bibinfo {volume} {13}},\ \bibinfo {pages}
  {1173} (\bibinfo {year} {2022})}\BibitemShut {NoStop}%
\bibitem [{\citenamefont {Yogo}\ \emph
  {et~al.}(2023{\natexlab{b}})\citenamefont {Yogo}, \citenamefont {Lan},
  \citenamefont {Arikawa}, \citenamefont {Abe}, \citenamefont {Mirfayzi},
  \citenamefont {Wei}, \citenamefont {Mori}, \citenamefont {Golovin},
  \citenamefont {Hayakawa}, \citenamefont {Iwata} \emph
  {et~al.}}]{yogo2023laser}%
  \BibitemOpen
  \bibfield  {author} {\bibinfo {author} {\bibfnamefont {A.}~\bibnamefont
  {Yogo}}, \bibinfo {author} {\bibfnamefont {Z.}~\bibnamefont {Lan}}, \bibinfo
  {author} {\bibfnamefont {Y.}~\bibnamefont {Arikawa}}, \bibinfo {author}
  {\bibfnamefont {Y.}~\bibnamefont {Abe}}, \bibinfo {author} {\bibfnamefont
  {S.}~\bibnamefont {Mirfayzi}}, \bibinfo {author} {\bibfnamefont
  {T.}~\bibnamefont {Wei}}, \bibinfo {author} {\bibfnamefont {T.}~\bibnamefont
  {Mori}}, \bibinfo {author} {\bibfnamefont {D.}~\bibnamefont {Golovin}},
  \bibinfo {author} {\bibfnamefont {T.}~\bibnamefont {Hayakawa}}, \bibinfo
  {author} {\bibfnamefont {N.}~\bibnamefont {Iwata}},  \emph {et~al.},\
  }\href@noop {} {\bibfield  {journal} {\bibinfo  {journal} {Physical Review
  X}\ }\textbf {\bibinfo {volume} {13}},\ \bibinfo {pages} {011011} (\bibinfo
  {year} {2023}{\natexlab{b}})}\BibitemShut {NoStop}%
\bibitem [{\citenamefont {Lancaster}\ \emph {et~al.}(2004)\citenamefont
  {Lancaster}, \citenamefont {Karsch}, \citenamefont {Habara}, \citenamefont
  {Beg}, \citenamefont {Clark}, \citenamefont {Freeman}, \citenamefont {Key},
  \citenamefont {King}, \citenamefont {Kodama}, \citenamefont {Krushelnick}
  \emph {et~al.}}]{lancaster2004characterization}%
  \BibitemOpen
  \bibfield  {author} {\bibinfo {author} {\bibfnamefont {K.}~\bibnamefont
  {Lancaster}}, \bibinfo {author} {\bibfnamefont {S.}~\bibnamefont {Karsch}},
  \bibinfo {author} {\bibfnamefont {H.}~\bibnamefont {Habara}}, \bibinfo
  {author} {\bibfnamefont {F.}~\bibnamefont {Beg}}, \bibinfo {author}
  {\bibfnamefont {E.}~\bibnamefont {Clark}}, \bibinfo {author} {\bibfnamefont
  {R.}~\bibnamefont {Freeman}}, \bibinfo {author} {\bibfnamefont
  {M.}~\bibnamefont {Key}}, \bibinfo {author} {\bibfnamefont {J.}~\bibnamefont
  {King}}, \bibinfo {author} {\bibfnamefont {R.}~\bibnamefont {Kodama}},
  \bibinfo {author} {\bibfnamefont {K.}~\bibnamefont {Krushelnick}},  \emph
  {et~al.},\ }\href@noop {} {\bibfield  {journal} {\bibinfo  {journal} {Physics
  of plasmas}\ }\textbf {\bibinfo {volume} {11}},\ \bibinfo {pages} {3404}
  (\bibinfo {year} {2004})}\BibitemShut {NoStop}%
\bibitem [{\citenamefont {Snavely}\ \emph {et~al.}(2000)\citenamefont
  {Snavely}, \citenamefont {Key}, \citenamefont {Hatchett}, \citenamefont
  {Cowan}, \citenamefont {Roth}, \citenamefont {Phillips}, \citenamefont
  {Stoyer}, \citenamefont {Henry}, \citenamefont {Sangster}, \citenamefont
  {Singh} \emph {et~al.}}]{snavely2000intense}%
  \BibitemOpen
  \bibfield  {author} {\bibinfo {author} {\bibfnamefont {R.}~\bibnamefont
  {Snavely}}, \bibinfo {author} {\bibfnamefont {M.}~\bibnamefont {Key}},
  \bibinfo {author} {\bibfnamefont {S.}~\bibnamefont {Hatchett}}, \bibinfo
  {author} {\bibfnamefont {T.}~\bibnamefont {Cowan}}, \bibinfo {author}
  {\bibfnamefont {M.}~\bibnamefont {Roth}}, \bibinfo {author} {\bibfnamefont
  {T.}~\bibnamefont {Phillips}}, \bibinfo {author} {\bibfnamefont
  {M.}~\bibnamefont {Stoyer}}, \bibinfo {author} {\bibfnamefont
  {E.}~\bibnamefont {Henry}}, \bibinfo {author} {\bibfnamefont
  {T.}~\bibnamefont {Sangster}}, \bibinfo {author} {\bibfnamefont
  {M.}~\bibnamefont {Singh}},  \emph {et~al.},\ }\href@noop {} {\bibfield
  {journal} {\bibinfo  {journal} {Physical review letters}\ }\textbf {\bibinfo
  {volume} {85}},\ \bibinfo {pages} {2945} (\bibinfo {year}
  {2000})}\BibitemShut {NoStop}%
\bibitem [{\citenamefont {Wilks}\ \emph {et~al.}(2001)\citenamefont {Wilks},
  \citenamefont {Langdon}, \citenamefont {Cowan}, \citenamefont {Roth},
  \citenamefont {Singh}, \citenamefont {Hatchett}, \citenamefont {Key},
  \citenamefont {Pennington}, \citenamefont {MacKinnon},\ and\ \citenamefont
  {Snavely}}]{wilks2001energetic}%
  \BibitemOpen
  \bibfield  {author} {\bibinfo {author} {\bibfnamefont {S.}~\bibnamefont
  {Wilks}}, \bibinfo {author} {\bibfnamefont {A.}~\bibnamefont {Langdon}},
  \bibinfo {author} {\bibfnamefont {T.}~\bibnamefont {Cowan}}, \bibinfo
  {author} {\bibfnamefont {M.}~\bibnamefont {Roth}}, \bibinfo {author}
  {\bibfnamefont {M.}~\bibnamefont {Singh}}, \bibinfo {author} {\bibfnamefont
  {S.}~\bibnamefont {Hatchett}}, \bibinfo {author} {\bibfnamefont
  {M.}~\bibnamefont {Key}}, \bibinfo {author} {\bibfnamefont {D.}~\bibnamefont
  {Pennington}}, \bibinfo {author} {\bibfnamefont {A.}~\bibnamefont
  {MacKinnon}}, \ and\ \bibinfo {author} {\bibfnamefont {R.}~\bibnamefont
  {Snavely}},\ }\href@noop {} {\bibfield  {journal} {\bibinfo  {journal}
  {Physics of plasmas}\ }\textbf {\bibinfo {volume} {8}},\ \bibinfo {pages}
  {542} (\bibinfo {year} {2001})}\BibitemShut {NoStop}%
\bibitem [{\citenamefont {Passoni}, \citenamefont {Bertagna},\ and\
  \citenamefont {Zani}(2010)}]{passoni2010target}%
  \BibitemOpen
  \bibfield  {author} {\bibinfo {author} {\bibfnamefont {M.}~\bibnamefont
  {Passoni}}, \bibinfo {author} {\bibfnamefont {L.}~\bibnamefont {Bertagna}}, \
  and\ \bibinfo {author} {\bibfnamefont {A.}~\bibnamefont {Zani}},\ }\href@noop
  {} {\bibfield  {journal} {\bibinfo  {journal} {New Journal of Physics}\
  }\textbf {\bibinfo {volume} {12}},\ \bibinfo {pages} {045012} (\bibinfo
  {year} {2010})}\BibitemShut {NoStop}%
\bibitem [{\citenamefont {McKenna}\ \emph {et~al.}(2013)\citenamefont
  {McKenna}, \citenamefont {Neely}, \citenamefont {Bingham},\ and\
  \citenamefont {Jaroszynski}}]{mckenna2013laser}%
  \BibitemOpen
  \bibfield  {author} {\bibinfo {author} {\bibfnamefont {P.}~\bibnamefont
  {McKenna}}, \bibinfo {author} {\bibfnamefont {D.}~\bibnamefont {Neely}},
  \bibinfo {author} {\bibfnamefont {R.}~\bibnamefont {Bingham}}, \ and\
  \bibinfo {author} {\bibfnamefont {D.}~\bibnamefont {Jaroszynski}},\
  }\href@noop {} {\emph {\bibinfo {title} {Laser-plasma interactions and
  applications}}}\ (\bibinfo  {publisher} {Springer},\ \bibinfo {year}
  {2013})\BibitemShut {NoStop}%
\bibitem [{\citenamefont {Kar}\ \emph {et~al.}(2016)\citenamefont {Kar},
  \citenamefont {Green}, \citenamefont {Ahmed}, \citenamefont {Alejo},
  \citenamefont {Robinson}, \citenamefont {Cerchez}, \citenamefont {Clarke},
  \citenamefont {Doria}, \citenamefont {Dorkings}, \citenamefont {Fernandez}
  \emph {et~al.}}]{kar2016beamed}%
  \BibitemOpen
  \bibfield  {author} {\bibinfo {author} {\bibfnamefont {S.}~\bibnamefont
  {Kar}}, \bibinfo {author} {\bibfnamefont {A.}~\bibnamefont {Green}}, \bibinfo
  {author} {\bibfnamefont {H.}~\bibnamefont {Ahmed}}, \bibinfo {author}
  {\bibfnamefont {A.}~\bibnamefont {Alejo}}, \bibinfo {author} {\bibfnamefont
  {A.}~\bibnamefont {Robinson}}, \bibinfo {author} {\bibfnamefont
  {M.}~\bibnamefont {Cerchez}}, \bibinfo {author} {\bibfnamefont
  {R.}~\bibnamefont {Clarke}}, \bibinfo {author} {\bibfnamefont
  {D.}~\bibnamefont {Doria}}, \bibinfo {author} {\bibfnamefont
  {S.}~\bibnamefont {Dorkings}}, \bibinfo {author} {\bibfnamefont
  {J.}~\bibnamefont {Fernandez}},  \emph {et~al.},\ }\href@noop {} {\bibfield
  {journal} {\bibinfo  {journal} {New Journal of Physics}\ }\textbf {\bibinfo
  {volume} {18}},\ \bibinfo {pages} {053002} (\bibinfo {year}
  {2016})}\BibitemShut {NoStop}%
\bibitem [{\citenamefont {Roth}\ \emph {et~al.}(2013)\citenamefont {Roth},
  \citenamefont {Jung}, \citenamefont {Falk}, \citenamefont {Guler},
  \citenamefont {Deppert}, \citenamefont {Devlin}, \citenamefont {Favalli},
  \citenamefont {Fernandez}, \citenamefont {Gautier}, \citenamefont {Geissel}
  \emph {et~al.}}]{roth2013bright}%
  \BibitemOpen
  \bibfield  {author} {\bibinfo {author} {\bibfnamefont {M.}~\bibnamefont
  {Roth}}, \bibinfo {author} {\bibfnamefont {D.}~\bibnamefont {Jung}}, \bibinfo
  {author} {\bibfnamefont {K.}~\bibnamefont {Falk}}, \bibinfo {author}
  {\bibfnamefont {N.}~\bibnamefont {Guler}}, \bibinfo {author} {\bibfnamefont
  {O.}~\bibnamefont {Deppert}}, \bibinfo {author} {\bibfnamefont
  {M.}~\bibnamefont {Devlin}}, \bibinfo {author} {\bibfnamefont
  {A.}~\bibnamefont {Favalli}}, \bibinfo {author} {\bibfnamefont
  {J.}~\bibnamefont {Fernandez}}, \bibinfo {author} {\bibfnamefont
  {D.}~\bibnamefont {Gautier}}, \bibinfo {author} {\bibfnamefont
  {M.}~\bibnamefont {Geissel}},  \emph {et~al.},\ }\href@noop {} {\bibfield
  {journal} {\bibinfo  {journal} {Physical review letters}\ }\textbf {\bibinfo
  {volume} {110}},\ \bibinfo {pages} {044802} (\bibinfo {year}
  {2013})}\BibitemShut {NoStop}%
\bibitem [{\citenamefont {Kleinschmidt}\ \emph {et~al.}(2018)\citenamefont
  {Kleinschmidt}, \citenamefont {Bagnoud}, \citenamefont {Deppert},
  \citenamefont {Favalli}, \citenamefont {Frydrych}, \citenamefont {Hornung},
  \citenamefont {Jahn}, \citenamefont {Schaumann}, \citenamefont {Tebartz},
  \citenamefont {Wagner}, \citenamefont {Wurden}, \citenamefont {Zielbauer},\
  and\ \citenamefont {Roth}}]{Kleinschmidt2018}%
  \BibitemOpen
  \bibfield  {author} {\bibinfo {author} {\bibfnamefont {A.}~\bibnamefont
  {Kleinschmidt}}, \bibinfo {author} {\bibfnamefont {V.}~\bibnamefont
  {Bagnoud}}, \bibinfo {author} {\bibfnamefont {O.}~\bibnamefont {Deppert}},
  \bibinfo {author} {\bibfnamefont {A.}~\bibnamefont {Favalli}}, \bibinfo
  {author} {\bibfnamefont {S.}~\bibnamefont {Frydrych}}, \bibinfo {author}
  {\bibfnamefont {J.}~\bibnamefont {Hornung}}, \bibinfo {author} {\bibfnamefont
  {D.}~\bibnamefont {Jahn}}, \bibinfo {author} {\bibfnamefont {G.}~\bibnamefont
  {Schaumann}}, \bibinfo {author} {\bibfnamefont {A.}~\bibnamefont {Tebartz}},
  \bibinfo {author} {\bibfnamefont {F.}~\bibnamefont {Wagner}}, \bibinfo
  {author} {\bibfnamefont {G.}~\bibnamefont {Wurden}}, \bibinfo {author}
  {\bibfnamefont {B.}~\bibnamefont {Zielbauer}}, \ and\ \bibinfo {author}
  {\bibfnamefont {M.}~\bibnamefont {Roth}},\ }\href {\doibase
  10.1063/1.5006613} {\bibfield  {journal} {\bibinfo  {journal} {Physics of
  Plasmas}\ }\textbf {\bibinfo {volume} {25}} (\bibinfo {year} {2018}),\
  10.1063/1.5006613}\BibitemShut {NoStop}%
\bibitem [{\citenamefont {Pomerantz}\ \emph {et~al.}(2014)\citenamefont
  {Pomerantz}, \citenamefont {Mccary}, \citenamefont {Meadows}, \citenamefont
  {Arefiev}, \citenamefont {Bernstein}, \citenamefont {Chester}, \citenamefont
  {Cortez}, \citenamefont {Donovan}, \citenamefont {Dyer}, \citenamefont {Gaul}
  \emph {et~al.}}]{pomerantz2014ultrashort}%
  \BibitemOpen
  \bibfield  {author} {\bibinfo {author} {\bibfnamefont {I.}~\bibnamefont
  {Pomerantz}}, \bibinfo {author} {\bibfnamefont {E.}~\bibnamefont {Mccary}},
  \bibinfo {author} {\bibfnamefont {A.~R.}\ \bibnamefont {Meadows}}, \bibinfo
  {author} {\bibfnamefont {A.}~\bibnamefont {Arefiev}}, \bibinfo {author}
  {\bibfnamefont {A.~C.}\ \bibnamefont {Bernstein}}, \bibinfo {author}
  {\bibfnamefont {C.}~\bibnamefont {Chester}}, \bibinfo {author} {\bibfnamefont
  {J.}~\bibnamefont {Cortez}}, \bibinfo {author} {\bibfnamefont {M.~E.}\
  \bibnamefont {Donovan}}, \bibinfo {author} {\bibfnamefont {G.}~\bibnamefont
  {Dyer}}, \bibinfo {author} {\bibfnamefont {E.~W.}\ \bibnamefont {Gaul}},
  \emph {et~al.},\ }\href@noop {} {\bibfield  {journal} {\bibinfo  {journal}
  {Physical review letters}\ }\textbf {\bibinfo {volume} {113}},\ \bibinfo
  {pages} {184801} (\bibinfo {year} {2014})}\BibitemShut {NoStop}%
\bibitem [{\citenamefont {Bradford}\ \emph {et~al.}(2018)\citenamefont
  {Bradford}, \citenamefont {Woolsey}, \citenamefont {Scott}, \citenamefont
  {Liao}, \citenamefont {Liu}, \citenamefont {Zhang}, \citenamefont {Zhu},
  \citenamefont {Armstrong}, \citenamefont {Astbury}, \citenamefont {Brenner}
  \emph {et~al.}}]{bradford2018emp}%
  \BibitemOpen
  \bibfield  {author} {\bibinfo {author} {\bibfnamefont {P.}~\bibnamefont
  {Bradford}}, \bibinfo {author} {\bibfnamefont {N.}~\bibnamefont {Woolsey}},
  \bibinfo {author} {\bibfnamefont {G.}~\bibnamefont {Scott}}, \bibinfo
  {author} {\bibfnamefont {G.}~\bibnamefont {Liao}}, \bibinfo {author}
  {\bibfnamefont {H.}~\bibnamefont {Liu}}, \bibinfo {author} {\bibfnamefont
  {Y.}~\bibnamefont {Zhang}}, \bibinfo {author} {\bibfnamefont
  {B.}~\bibnamefont {Zhu}}, \bibinfo {author} {\bibfnamefont {C.}~\bibnamefont
  {Armstrong}}, \bibinfo {author} {\bibfnamefont {S.}~\bibnamefont {Astbury}},
  \bibinfo {author} {\bibfnamefont {C.}~\bibnamefont {Brenner}},  \emph
  {et~al.},\ }\href@noop {} {\bibfield  {journal} {\bibinfo  {journal} {High
  Power Laser Science and Engineering}\ }\textbf {\bibinfo {volume} {6}},\
  \bibinfo {pages} {e21} (\bibinfo {year} {2018})}\BibitemShut {NoStop}%
\bibitem [{\citenamefont {Seimetz}\ \emph {et~al.}(2020)\citenamefont
  {Seimetz}, \citenamefont {Bellido}, \citenamefont {Mur}, \citenamefont
  {Lera}, \citenamefont {Ruiz-de~la Cruz}, \citenamefont {S{\'a}nchez},
  \citenamefont {Zaffino}, \citenamefont {Benlliure}, \citenamefont {Ruiz},
  \citenamefont {Roso} \emph {et~al.}}]{seimetz2020electromagnetic}%
  \BibitemOpen
  \bibfield  {author} {\bibinfo {author} {\bibfnamefont {M.}~\bibnamefont
  {Seimetz}}, \bibinfo {author} {\bibfnamefont {P.}~\bibnamefont {Bellido}},
  \bibinfo {author} {\bibfnamefont {P.}~\bibnamefont {Mur}}, \bibinfo {author}
  {\bibfnamefont {R.}~\bibnamefont {Lera}}, \bibinfo {author} {\bibfnamefont
  {A.}~\bibnamefont {Ruiz-de~la Cruz}}, \bibinfo {author} {\bibfnamefont
  {I.}~\bibnamefont {S{\'a}nchez}}, \bibinfo {author} {\bibfnamefont
  {R.}~\bibnamefont {Zaffino}}, \bibinfo {author} {\bibfnamefont
  {J.}~\bibnamefont {Benlliure}}, \bibinfo {author} {\bibfnamefont
  {C.}~\bibnamefont {Ruiz}}, \bibinfo {author} {\bibfnamefont {L.}~\bibnamefont
  {Roso}},  \emph {et~al.},\ }\href@noop {} {\bibfield  {journal} {\bibinfo
  {journal} {Plasma Physics and Controlled Fusion}\ }\textbf {\bibinfo {volume}
  {62}},\ \bibinfo {pages} {115008} (\bibinfo {year} {2020})}\BibitemShut
  {NoStop}%
\bibitem [{\citenamefont {Ing}\ and\ \citenamefont {Birnboim}(1984)}]{Ing1984}%
  \BibitemOpen
  \bibfield  {author} {\bibinfo {author} {\bibfnamefont {H.}~\bibnamefont
  {Ing}}\ and\ \bibinfo {author} {\bibfnamefont {H.}~\bibnamefont {Birnboim}},\
  }\href {\doibase 10.1016/0735-245x(84)90106-6} {\bibfield  {journal}
  {\bibinfo  {journal} {Nuclear Tracks and Radiation Measurements (1982)}\
  }\textbf {\bibinfo {volume} {8}},\ \bibinfo {pages} {285} (\bibinfo {year}
  {1984})}\BibitemShut {NoStop}%
\bibitem [{\citenamefont {Ing}, \citenamefont {Noulty},\ and\ \citenamefont
  {McLean}(1997)}]{Ing1997}%
  \BibitemOpen
  \bibfield  {author} {\bibinfo {author} {\bibfnamefont {H.}~\bibnamefont
  {Ing}}, \bibinfo {author} {\bibfnamefont {R.}~\bibnamefont {Noulty}}, \ and\
  \bibinfo {author} {\bibfnamefont {T.}~\bibnamefont {McLean}},\ }\href
  {\doibase 10.1016/s1350-4487(96)00156-4} {\bibfield  {journal} {\bibinfo
  {journal} {Radiation Measurements}\ }\textbf {\bibinfo {volume} {27}},\
  \bibinfo {pages} {1} (\bibinfo {year} {1997})}\BibitemShut {NoStop}%
\bibitem [{\citenamefont {{Bubble Technology Industries}}(2023)}]{BTI}%
  \BibitemOpen
  \bibfield  {author} {\bibinfo {author} {\bibnamefont {{Bubble Technology
  Industries}}},\ }\href {http://www.bubbletech.ca} {} (\bibinfo {year}
  {2023})\BibitemShut {NoStop}%
\bibitem [{\citenamefont {Smith}\ \emph {et~al.}(2014)\citenamefont {Smith},
  \citenamefont {Andrews}, \citenamefont {Ing},\ and\ \citenamefont
  {Koslowsky}}]{Smith2014}%
  \BibitemOpen
  \bibfield  {author} {\bibinfo {author} {\bibfnamefont {M.~B.}\ \bibnamefont
  {Smith}}, \bibinfo {author} {\bibfnamefont {H.~R.}\ \bibnamefont {Andrews}},
  \bibinfo {author} {\bibfnamefont {H.}~\bibnamefont {Ing}}, \ and\ \bibinfo
  {author} {\bibfnamefont {M.~R.}\ \bibnamefont {Koslowsky}},\ }\href {\doibase
  10.1093/rpd/ncu288} {\bibfield  {journal} {\bibinfo  {journal} {Radiation
  Protection Dosimetry}\ }\textbf {\bibinfo {volume} {164}},\ \bibinfo {pages}
  {203} (\bibinfo {year} {2014})}\BibitemShut {NoStop}%
\bibitem [{\citenamefont {Mirfayzi}\ \emph {et~al.}(2015)\citenamefont
  {Mirfayzi}, \citenamefont {Kar}, \citenamefont {Ahmed}, \citenamefont
  {Krygier}, \citenamefont {Green}, \citenamefont {Alejo}, \citenamefont
  {Clarke}, \citenamefont {Freeman}, \citenamefont {Fuchs}, \citenamefont
  {Jung} \emph {et~al.}}]{mirfayzi2015calibration}%
  \BibitemOpen
  \bibfield  {author} {\bibinfo {author} {\bibfnamefont {S.}~\bibnamefont
  {Mirfayzi}}, \bibinfo {author} {\bibfnamefont {S.}~\bibnamefont {Kar}},
  \bibinfo {author} {\bibfnamefont {H.}~\bibnamefont {Ahmed}}, \bibinfo
  {author} {\bibfnamefont {A.}~\bibnamefont {Krygier}}, \bibinfo {author}
  {\bibfnamefont {A.}~\bibnamefont {Green}}, \bibinfo {author} {\bibfnamefont
  {A.}~\bibnamefont {Alejo}}, \bibinfo {author} {\bibfnamefont
  {R.}~\bibnamefont {Clarke}}, \bibinfo {author} {\bibfnamefont
  {R.}~\bibnamefont {Freeman}}, \bibinfo {author} {\bibfnamefont
  {J.}~\bibnamefont {Fuchs}}, \bibinfo {author} {\bibfnamefont
  {D.}~\bibnamefont {Jung}},  \emph {et~al.},\ }\href@noop {} {\bibfield
  {journal} {\bibinfo  {journal} {Review of Scientific Instruments}\ }\textbf
  {\bibinfo {volume} {86}} (\bibinfo {year} {2015})}\BibitemShut {NoStop}%
\bibitem [{\citenamefont {Treffert}(2023)}]{trefferthigh}%
  \BibitemOpen
  \bibfield  {author} {\bibinfo {author} {\bibfnamefont {F.}~\bibnamefont
  {Treffert}},\ }\emph {\bibinfo {title} {High Repetition-Rate Laser-Driven
  Particle Generation}},\ \href {https://tuprints.ulb.tu-darmstadt.de/23183/}
  {Ph.D. thesis},\ \bibinfo  {school} {Technische Universit{\"a}t Darmstadt}
  (\bibinfo {year} {2023})\BibitemShut {NoStop}%
\bibitem [{\citenamefont {Jung}\ \emph {et~al.}(2013)\citenamefont {Jung},
  \citenamefont {Falk}, \citenamefont {Guler}, \citenamefont {Deppert},
  \citenamefont {Devlin}, \citenamefont {Favalli}, \citenamefont {Fernandez},
  \citenamefont {Gautier}, \citenamefont {Geissel}, \citenamefont {Haight}
  \emph {et~al.}}]{jung2013characterization}%
  \BibitemOpen
  \bibfield  {author} {\bibinfo {author} {\bibfnamefont {D.}~\bibnamefont
  {Jung}}, \bibinfo {author} {\bibfnamefont {K.}~\bibnamefont {Falk}}, \bibinfo
  {author} {\bibfnamefont {N.}~\bibnamefont {Guler}}, \bibinfo {author}
  {\bibfnamefont {O.}~\bibnamefont {Deppert}}, \bibinfo {author} {\bibfnamefont
  {M.}~\bibnamefont {Devlin}}, \bibinfo {author} {\bibfnamefont
  {A.}~\bibnamefont {Favalli}}, \bibinfo {author} {\bibfnamefont
  {J.}~\bibnamefont {Fernandez}}, \bibinfo {author} {\bibfnamefont
  {D.}~\bibnamefont {Gautier}}, \bibinfo {author} {\bibfnamefont
  {M.}~\bibnamefont {Geissel}}, \bibinfo {author} {\bibfnamefont
  {R.}~\bibnamefont {Haight}},  \emph {et~al.},\ }\href@noop {} {\bibfield
  {journal} {\bibinfo  {journal} {Physics of Plasmas}\ }\textbf {\bibinfo
  {volume} {20}} (\bibinfo {year} {2013})}\BibitemShut {NoStop}%
\bibitem [{\citenamefont {Ziegler}\ \emph {et~al.}(2021)\citenamefont
  {Ziegler}, \citenamefont {Albach}, \citenamefont {Bernert}, \citenamefont
  {Bock}, \citenamefont {Brack}, \citenamefont {Cowan}, \citenamefont {Dover},
  \citenamefont {Garten}, \citenamefont {Gaus}, \citenamefont {Gebhardt},
  \citenamefont {Goethel}, \citenamefont {Helbig}, \citenamefont {Irman},
  \citenamefont {Kiriyama}, \citenamefont {Kluge}, \citenamefont {Kon},
  \citenamefont {Kraft}, \citenamefont {Kroll}, \citenamefont {Loeser},
  \citenamefont {Metzkes-Ng}, \citenamefont {Nishiuchi}, \citenamefont
  {Obst-Huebl}, \citenamefont {Püschel}, \citenamefont {Rehwald},
  \citenamefont {Schlenvoigt}, \citenamefont {Schramm},\ and\ \citenamefont
  {Zeil}}]{Ziegler_2021}%
  \BibitemOpen
  \bibfield  {author} {\bibinfo {author} {\bibfnamefont {T.}~\bibnamefont
  {Ziegler}}, \bibinfo {author} {\bibfnamefont {D.}~\bibnamefont {Albach}},
  \bibinfo {author} {\bibfnamefont {C.}~\bibnamefont {Bernert}}, \bibinfo
  {author} {\bibfnamefont {S.}~\bibnamefont {Bock}}, \bibinfo {author}
  {\bibfnamefont {F.-E.}\ \bibnamefont {Brack}}, \bibinfo {author}
  {\bibfnamefont {T.~E.}\ \bibnamefont {Cowan}}, \bibinfo {author}
  {\bibfnamefont {N.~P.}\ \bibnamefont {Dover}}, \bibinfo {author}
  {\bibfnamefont {M.}~\bibnamefont {Garten}}, \bibinfo {author} {\bibfnamefont
  {L.}~\bibnamefont {Gaus}}, \bibinfo {author} {\bibfnamefont {R.}~\bibnamefont
  {Gebhardt}}, \bibinfo {author} {\bibfnamefont {I.}~\bibnamefont {Goethel}},
  \bibinfo {author} {\bibfnamefont {U.}~\bibnamefont {Helbig}}, \bibinfo
  {author} {\bibfnamefont {A.}~\bibnamefont {Irman}}, \bibinfo {author}
  {\bibfnamefont {H.}~\bibnamefont {Kiriyama}}, \bibinfo {author}
  {\bibfnamefont {T.}~\bibnamefont {Kluge}}, \bibinfo {author} {\bibfnamefont
  {A.}~\bibnamefont {Kon}}, \bibinfo {author} {\bibfnamefont {S.}~\bibnamefont
  {Kraft}}, \bibinfo {author} {\bibfnamefont {F.}~\bibnamefont {Kroll}},
  \bibinfo {author} {\bibfnamefont {M.}~\bibnamefont {Loeser}}, \bibinfo
  {author} {\bibfnamefont {J.}~\bibnamefont {Metzkes-Ng}}, \bibinfo {author}
  {\bibfnamefont {M.}~\bibnamefont {Nishiuchi}}, \bibinfo {author}
  {\bibfnamefont {L.}~\bibnamefont {Obst-Huebl}}, \bibinfo {author}
  {\bibfnamefont {T.}~\bibnamefont {Püschel}}, \bibinfo {author}
  {\bibfnamefont {M.}~\bibnamefont {Rehwald}}, \bibinfo {author} {\bibfnamefont
  {H.-P.}\ \bibnamefont {Schlenvoigt}}, \bibinfo {author} {\bibfnamefont
  {U.}~\bibnamefont {Schramm}}, \ and\ \bibinfo {author} {\bibfnamefont
  {K.}~\bibnamefont {Zeil}},\ }\href {\doibase 10.1038/s41598-021-86547-x}
  {\bibfield  {journal} {\bibinfo  {journal} {Scientific Reports}\ }\textbf
  {\bibinfo {volume} {11}} (\bibinfo {year} {2021}),\
  10.1038/s41598-021-86547-x}\BibitemShut {NoStop}%
\bibitem [{\citenamefont {Sato}\ \emph {et~al.}(2024)\citenamefont {Sato},
  \citenamefont {Iwamoto}, \citenamefont {Hashimoto}, \citenamefont {Ogawa},
  \citenamefont {Furuta}, \citenamefont {Abe}, \citenamefont {Kai},
  \citenamefont {Matsuya}, \citenamefont {Matsuda}, \citenamefont {Hirata}
  \emph {et~al.}}]{sato2024recent}%
  \BibitemOpen
  \bibfield  {author} {\bibinfo {author} {\bibfnamefont {T.}~\bibnamefont
  {Sato}}, \bibinfo {author} {\bibfnamefont {Y.}~\bibnamefont {Iwamoto}},
  \bibinfo {author} {\bibfnamefont {S.}~\bibnamefont {Hashimoto}}, \bibinfo
  {author} {\bibfnamefont {T.}~\bibnamefont {Ogawa}}, \bibinfo {author}
  {\bibfnamefont {T.}~\bibnamefont {Furuta}}, \bibinfo {author} {\bibfnamefont
  {S.-I.}\ \bibnamefont {Abe}}, \bibinfo {author} {\bibfnamefont
  {T.}~\bibnamefont {Kai}}, \bibinfo {author} {\bibfnamefont {Y.}~\bibnamefont
  {Matsuya}}, \bibinfo {author} {\bibfnamefont {N.}~\bibnamefont {Matsuda}},
  \bibinfo {author} {\bibfnamefont {Y.}~\bibnamefont {Hirata}},  \emph
  {et~al.},\ }\href@noop {} {\bibfield  {journal} {\bibinfo  {journal} {Journal
  of Nuclear Science and Technology}\ }\textbf {\bibinfo {volume} {61}},\
  \bibinfo {pages} {127} (\bibinfo {year} {2024})}\BibitemShut {NoStop}%
\bibitem [{\citenamefont {Iwamoto}\ \emph {et~al.}(2022)\citenamefont
  {Iwamoto}, \citenamefont {Hashimoto}, \citenamefont {Sato}, \citenamefont
  {Matsuda}, \citenamefont {Kunieda}, \citenamefont {{\c{C}}elik},
  \citenamefont {Furutachi},\ and\ \citenamefont
  {Niita}}]{iwamoto2022benchmark}%
  \BibitemOpen
  \bibfield  {author} {\bibinfo {author} {\bibfnamefont {Y.}~\bibnamefont
  {Iwamoto}}, \bibinfo {author} {\bibfnamefont {S.}~\bibnamefont {Hashimoto}},
  \bibinfo {author} {\bibfnamefont {T.}~\bibnamefont {Sato}}, \bibinfo {author}
  {\bibfnamefont {N.}~\bibnamefont {Matsuda}}, \bibinfo {author} {\bibfnamefont
  {S.}~\bibnamefont {Kunieda}}, \bibinfo {author} {\bibfnamefont
  {Y.}~\bibnamefont {{\c{C}}elik}}, \bibinfo {author} {\bibfnamefont
  {N.}~\bibnamefont {Furutachi}}, \ and\ \bibinfo {author} {\bibfnamefont
  {K.}~\bibnamefont {Niita}},\ }\href@noop {} {\bibfield  {journal} {\bibinfo
  {journal} {Journal of Nuclear Science and Technology}\ }\textbf {\bibinfo
  {volume} {59}},\ \bibinfo {pages} {665} (\bibinfo {year} {2022})}\BibitemShut
  {NoStop}%
\bibitem [{\citenamefont {Iwamoto}\ \emph {et~al.}(2017)\citenamefont
  {Iwamoto}, \citenamefont {Sato}, \citenamefont {Hashimoto}, \citenamefont
  {Ogawa}, \citenamefont {Furuta}, \citenamefont {Abe}, \citenamefont {Kai},
  \citenamefont {Matsuda}, \citenamefont {Hosoyamada},\ and\ \citenamefont
  {Niita}}]{iwamoto2017benchmark}%
  \BibitemOpen
  \bibfield  {author} {\bibinfo {author} {\bibfnamefont {Y.}~\bibnamefont
  {Iwamoto}}, \bibinfo {author} {\bibfnamefont {T.}~\bibnamefont {Sato}},
  \bibinfo {author} {\bibfnamefont {S.}~\bibnamefont {Hashimoto}}, \bibinfo
  {author} {\bibfnamefont {T.}~\bibnamefont {Ogawa}}, \bibinfo {author}
  {\bibfnamefont {T.}~\bibnamefont {Furuta}}, \bibinfo {author} {\bibfnamefont
  {S.-i.}\ \bibnamefont {Abe}}, \bibinfo {author} {\bibfnamefont
  {T.}~\bibnamefont {Kai}}, \bibinfo {author} {\bibfnamefont {N.}~\bibnamefont
  {Matsuda}}, \bibinfo {author} {\bibfnamefont {R.}~\bibnamefont {Hosoyamada}},
  \ and\ \bibinfo {author} {\bibfnamefont {K.}~\bibnamefont {Niita}},\
  }\href@noop {} {\bibfield  {journal} {\bibinfo  {journal} {Journal of Nuclear
  Science and Technology}\ }\textbf {\bibinfo {volume} {54}},\ \bibinfo {pages}
  {617} (\bibinfo {year} {2017})}\BibitemShut {NoStop}%
\bibitem [{\citenamefont {Iwamoto}, \citenamefont {Tsuda},\ and\ \citenamefont
  {Ogawa}(2023)}]{iwamoto2023benchmark}%
  \BibitemOpen
  \bibfield  {author} {\bibinfo {author} {\bibfnamefont {Y.}~\bibnamefont
  {Iwamoto}}, \bibinfo {author} {\bibfnamefont {S.}~\bibnamefont {Tsuda}}, \
  and\ \bibinfo {author} {\bibfnamefont {T.}~\bibnamefont {Ogawa}},\
  }\href@noop {} {\bibfield  {journal} {\bibinfo  {journal} {Frontiers in
  Energy Research}\ }\textbf {\bibinfo {volume} {11}},\ \bibinfo {pages}
  {1085264} (\bibinfo {year} {2023})}\BibitemShut {NoStop}%
\bibitem [{\citenamefont {Saves}\ \emph {et~al.}(2024)\citenamefont {Saves},
  \citenamefont {Lafage}, \citenamefont {Bartoli}, \citenamefont {Diouane},
  \citenamefont {Bussemaker}, \citenamefont {Lefebvre}, \citenamefont {Hwang},
  \citenamefont {Morlier},\ and\ \citenamefont {Martins}}]{Saves2024}%
  \BibitemOpen
  \bibfield  {author} {\bibinfo {author} {\bibfnamefont {P.}~\bibnamefont
  {Saves}}, \bibinfo {author} {\bibfnamefont {R.}~\bibnamefont {Lafage}},
  \bibinfo {author} {\bibfnamefont {N.}~\bibnamefont {Bartoli}}, \bibinfo
  {author} {\bibfnamefont {Y.}~\bibnamefont {Diouane}}, \bibinfo {author}
  {\bibfnamefont {J.}~\bibnamefont {Bussemaker}}, \bibinfo {author}
  {\bibfnamefont {T.}~\bibnamefont {Lefebvre}}, \bibinfo {author}
  {\bibfnamefont {J.~T.}\ \bibnamefont {Hwang}}, \bibinfo {author}
  {\bibfnamefont {J.}~\bibnamefont {Morlier}}, \ and\ \bibinfo {author}
  {\bibfnamefont {J.~R.}\ \bibnamefont {Martins}},\ }\href {\doibase
  10.1016/j.advengsoft.2023.103571} {\bibfield  {journal} {\bibinfo  {journal}
  {Advances in Engineering Software}\ }\textbf {\bibinfo {volume} {188}},\
  \bibinfo {pages} {103571} (\bibinfo {year} {2024})}\BibitemShut {NoStop}%
\bibitem [{\citenamefont {Rasmussen}\ and\ \citenamefont
  {Williams}(2006)}]{RasmussenW06}%
  \BibitemOpen
  \bibfield  {author} {\bibinfo {author} {\bibfnamefont {C.~E.}\ \bibnamefont
  {Rasmussen}}\ and\ \bibinfo {author} {\bibfnamefont {C.~K.~I.}\ \bibnamefont
  {Williams}},\ }\href@noop {} {\emph {\bibinfo {title} {Gaussian processes for
  machine learning.}}},\ Adaptive computation and machine learning\ (\bibinfo
  {publisher} {MIT Press},\ \bibinfo {year} {2006})\ pp.\ \bibinfo {pages}
  {I--XVIII, 1--248}\BibitemShut {NoStop}%
\bibitem [{\citenamefont {Schmitz}, \citenamefont {Metternich},\ and\
  \citenamefont {Boine-Frankenheim}(2022)}]{Schmitz_2022}%
  \BibitemOpen
  \bibfield  {author} {\bibinfo {author} {\bibfnamefont {B.}~\bibnamefont
  {Schmitz}}, \bibinfo {author} {\bibfnamefont {M.}~\bibnamefont {Metternich}},
  \ and\ \bibinfo {author} {\bibfnamefont {O.}~\bibnamefont
  {Boine-Frankenheim}},\ }\href {\doibase 10.1063/5.0094105} {\bibfield
  {journal} {\bibinfo  {journal} {Review of Scientific Instruments}\ }\textbf
  {\bibinfo {volume} {93}} (\bibinfo {year} {2022}),\
  10.1063/5.0094105}\BibitemShut {NoStop}%
\bibitem [{\citenamefont {Nürnberg}\ \emph {et~al.}(2009)\citenamefont
  {Nürnberg}, \citenamefont {Schollmeier}, \citenamefont {Brambrink},
  \citenamefont {Blažević}, \citenamefont {Carroll}, \citenamefont {Flippo},
  \citenamefont {Gautier}, \citenamefont {Geißel}, \citenamefont {Harres},
  \citenamefont {Hegelich}, \citenamefont {Lundh}, \citenamefont {Markey},
  \citenamefont {McKenna}, \citenamefont {Neely}, \citenamefont {Schreiber},\
  and\ \citenamefont {Roth}}]{Nuernberg_2009}%
  \BibitemOpen
  \bibfield  {author} {\bibinfo {author} {\bibfnamefont {F.}~\bibnamefont
  {Nürnberg}}, \bibinfo {author} {\bibfnamefont {M.}~\bibnamefont
  {Schollmeier}}, \bibinfo {author} {\bibfnamefont {E.}~\bibnamefont
  {Brambrink}}, \bibinfo {author} {\bibfnamefont {A.}~\bibnamefont
  {Blažević}}, \bibinfo {author} {\bibfnamefont {D.~C.}\ \bibnamefont
  {Carroll}}, \bibinfo {author} {\bibfnamefont {K.}~\bibnamefont {Flippo}},
  \bibinfo {author} {\bibfnamefont {D.~C.}\ \bibnamefont {Gautier}}, \bibinfo
  {author} {\bibfnamefont {M.}~\bibnamefont {Geißel}}, \bibinfo {author}
  {\bibfnamefont {K.}~\bibnamefont {Harres}}, \bibinfo {author} {\bibfnamefont
  {B.~M.}\ \bibnamefont {Hegelich}}, \bibinfo {author} {\bibfnamefont
  {O.}~\bibnamefont {Lundh}}, \bibinfo {author} {\bibfnamefont
  {K.}~\bibnamefont {Markey}}, \bibinfo {author} {\bibfnamefont
  {P.}~\bibnamefont {McKenna}}, \bibinfo {author} {\bibfnamefont
  {D.}~\bibnamefont {Neely}}, \bibinfo {author} {\bibfnamefont
  {J.}~\bibnamefont {Schreiber}}, \ and\ \bibinfo {author} {\bibfnamefont
  {M.}~\bibnamefont {Roth}},\ }\href {\doibase 10.1063/1.3086424} {\bibfield
  {journal} {\bibinfo  {journal} {Review of Scientific Instruments}\ }\textbf
  {\bibinfo {volume} {80}} (\bibinfo {year} {2009}),\
  10.1063/1.3086424}\BibitemShut {NoStop}%
\bibitem [{\citenamefont {Olsher}\ \emph {et~al.}(2007)\citenamefont {Olsher},
  \citenamefont {McLean}, \citenamefont {Mallett}, \citenamefont {Romero},
  \citenamefont {Devine},\ and\ \citenamefont {Hoffman}}]{olsher2007high}%
  \BibitemOpen
  \bibfield  {author} {\bibinfo {author} {\bibfnamefont {R.~H.}\ \bibnamefont
  {Olsher}}, \bibinfo {author} {\bibfnamefont {T.~D.}\ \bibnamefont {McLean}},
  \bibinfo {author} {\bibfnamefont {M.~W.}\ \bibnamefont {Mallett}}, \bibinfo
  {author} {\bibfnamefont {L.~L.}\ \bibnamefont {Romero}}, \bibinfo {author}
  {\bibfnamefont {R.~T.}\ \bibnamefont {Devine}}, \ and\ \bibinfo {author}
  {\bibfnamefont {J.~M.}\ \bibnamefont {Hoffman}},\ }\href@noop {} {\bibfield
  {journal} {\bibinfo  {journal} {Radiation protection dosimetry}\ }\textbf
  {\bibinfo {volume} {126}},\ \bibinfo {pages} {326} (\bibinfo {year}
  {2007})}\BibitemShut {NoStop}%
\bibitem [{\citenamefont {Smecka}\ and\ \citenamefont
  {Hajek}(2007)}]{smecka2007neutronendosimetrie}%
  \BibitemOpen
  \bibfield  {author} {\bibinfo {author} {\bibfnamefont {F.}~\bibnamefont
  {Smecka}}\ and\ \bibinfo {author} {\bibfnamefont {M.}~\bibnamefont {Hajek}},\
  }\href@noop {} {\bibfield  {journal} {\bibinfo  {journal} {AIAU Paper}\ }
  (\bibinfo {year} {2007})}\BibitemShut {NoStop}%
\bibitem [{\citenamefont {Bubble Technology~Industries}(2024)}]{bti2024reader}%
  \BibitemOpen
  \bibfield  {author} {\bibinfo {author} {\bibfnamefont {B.}~\bibnamefont
  {Bubble Technology~Industries}},\ }\href@noop {} {\enquote {\bibinfo {title}
  {Bdr-iii – bubble detector reader iii},}\ }\bibinfo {howpublished}
  {https://bubbletech.ca/product/bdr-iii-bubble-detector-reader-iii/} (\bibinfo
  {year} {2024}),\ \bibinfo {note} {accessed on Feb 05, 2024}\BibitemShut
  {NoStop}%
\bibitem [{\citenamefont {Buckner}, \citenamefont {Noulty},\ and\ \citenamefont
  {Cousins}(1994)}]{Buckner1994}%
  \BibitemOpen
  \bibfield  {author} {\bibinfo {author} {\bibfnamefont {M.}~\bibnamefont
  {Buckner}}, \bibinfo {author} {\bibfnamefont {R.}~\bibnamefont {Noulty}}, \
  and\ \bibinfo {author} {\bibfnamefont {T.}~\bibnamefont {Cousins}},\ }\href
  {\doibase 10.1093/oxfordjournals.rpd.a082371} {\bibfield  {journal} {\bibinfo
   {journal} {Radiation Protection Dosimetry}\ }\textbf {\bibinfo {volume}
  {55}},\ \bibinfo {pages} {23} (\bibinfo {year} {1994})}\BibitemShut {NoStop}%
\bibitem [{\citenamefont {Ing}(2001)}]{Ing2001}%
  \BibitemOpen
  \bibfield  {author} {\bibinfo {author} {\bibfnamefont {H.}~\bibnamefont
  {Ing}},\ }\href {\doibase 10.1016/s1350-4487(00)00154-2} {\bibfield
  {journal} {\bibinfo  {journal} {Radiation Measurements}\ }\textbf {\bibinfo
  {volume} {33}},\ \bibinfo {pages} {275} (\bibinfo {year} {2001})}\BibitemShut
  {NoStop}%
\end{thebibliography}%

\end{document}